\documentclass{aa}  
\usepackage{graphicx}
\usepackage{txfonts}
\usepackage{stfloats}
\usepackage{tabularx}
\usepackage[table,xcdraw]{xcolor}
\usepackage{url}
\usepackage{xcolor}
\usepackage{natbib}
\bibpunct{(}{)}{;}{a}{}{,} 
\usepackage{array}
\usepackage[colorlinks=true, linkcolor=blue, citecolor=blue, urlcolor=blue]{hyperref}
\usepackage{gensymb}
\usepackage{subfigure}
\usepackage{amsmath}
\usepackage{orcidlink}
\usepackage{comment}
\usepackage{svg}
\usepackage{placeins} 
\usepackage{subfigure}

\begin{document} 

   \title{Cloudy solutions for the clear skies of WASP-80b: 3D cloud feedback on the atmosphere and spectra of a warm Jupiter}

   \author{Nishil Mehta\inst{1}\thanks{nishil.mehta@oca.eu}
          \and
          Vivien Parmentier\inst{1}
          \and
          Xianyu Tan\inst{2}
          \and
          Elspeth K. H. Lee\inst{3}
          \and
          Tristan Guillot\inst{1}
          \and
          Lindsey S. Wiser\inst{4}
          \and
          Taylor J. Bell\inst{5}
          \and
          Everett Schlawin\inst{6}
          \and
          Kenneth Arnold\inst{7}
          \and
          Sagnick Mukherjee\inst{8,9}
          \and
          Thomas P. Greene\inst{10}
          \and
          Thomas G. Beatty\inst{7}
          \and
          Luis Welbanks\inst{11}
          \and
          Michael R. Line\inst{11}
          \and
          Matthew M. Murphy\inst{12}
          \and
          Jonathan J. Fortney\inst{13}
          \and
          Kazumasa Ohno\inst{14}
          }

   \institute{
   Université de la Côte d'Azur, Observatoire de la Côte d'Azur, CNRS, Laboratoire Lagrange, France 
        \and
            Tsung-Dao Lee Institute \& School of Physics and Astronomy, Shanghai Jiao Tong University, Shanghai 201210, People’s Republic of China
        \and
            Center for Space and Habitability, University of Bern, Gesellschaftsstrasse 6, CH-3012 Bern, Switzerland
        \and
            Johns Hopkins Applied Physics Laboratory,
Laurel, MD 20723, USA
        \and
            AURA for the European Space Agency (ESA), Space Telescope Science Institute, 3700 San Martin Drive, Baltimore, MD 21218, USA
        \and
            Steward Observatory, 933 North Cherry Avenue, Tucson, AZ 85721, USA
        \and
            Department of Astronomy, University of Wisconsin–Madison, Madison, WI 53703, USA
        \and
            Department of Astronomy and Astrophysics, University of California at Santa Cruz, Santa Cruz, CA 95064, USA
        \and
            Department of Physics and Astronomy, Johns Hopkins University, Baltimore, MD, USA
        \and
            IPAC, MC 100-22, California Institute of Technology, 1200 E. California Blvd., Pasadena, CA, 91125, USA
        \and
            School of Earth and Space Exploration, Arizona State University, Tempe, AZ, USA
        \and
            Department of Physics and Astronomy, Michigan State University, East Lansing, MI 48824, USA
        \and
            Department of Astronomy and Astrophysics, University of California, Santa Cruz, CA, USA
        \and
            Division of Science, National Astronomical Observatory of Japan, 2-12-1 Osawa, Mitaka-shi 1818588 Tokyo, Japan
            }

   \date{}

  \abstract
   {Close-in warm Jupiters orbiting M dwarf stars are expected to exhibit diverse atmospheric chemistry, with clouds playing a key role in shaping their albedo, heat distribution, and spectral properties.}
   {We study WASP-80b, a warm Jupiter orbiting an M dwarf star, using the latest JWST panchromatic emission and transmission spectra to comprehensively characterise its atmosphere, including cloud coverage, chemical composition, and particle sizes, and compare the observations with predictions from the general circulation models (GCMs).}
   {We used a GCM, ADAM (ADvanced Atmospheric MITgcm, formerly known as SPARC/MITgcm), combined with the latest JWST data to study the atmosphere of WASP-80b. A cloud module with radiatively active, tracer-based clouds was integrated with the GCM to study the effects on the atmosphere and the spectrum.}
   {
    We find that the emission and transmission spectra of WASP-80b are only compatible with cloudless atmospheres or with clouds composed of sufficiently large particles, namely Na$_2$S ($\geq 10~\mu$m), KCl ($\geq 1~\mu$m), and MgSiO$_3$ ($\geq 5~\mu$m). For these large-particle cloud cases, efficient gravitational settling confines the clouds to deeper atmospheric layers, resulting in weak spectral signatures. Smaller particles are ruled out due to their strong radiative feedback on the atmospheric structure.}
   {Overall, our results suggest that WASP-80b’s atmosphere is either effectively cloud-free or contains clouds composed of large, settled particles whose opacity has little impact on the observable atmosphere. This underscores the importance of particle size and vertical cloud distribution in interpreting exoplanet spectra. Future observations at shorter wavelengths may help distinguish between large-particle cloud scenarios and a truly cloudless atmosphere.
  }

    \titlerunning{GCM study of cloud feedback in WASP-80b}
   \maketitle

\section{Introduction}
\label{section: introduction}

JWST allows us to discover the diversity of exoplanet atmospheres. While past observatories have mainly focused on characterising hot Jupiters due to their favourable observational metrics, JWST opens up new opportunities to study the population of Jovian planets with temperatures cooler than 1000 K, also known as warm Jupiters. Whereas numerous studies have pointed out the role of 3D atmospheric transport (\cite{showman2011}, detailed review in \cite{showman2020}) and cloud formation in hot Jupiter atmospheres \citep{charnay2015, parmentier2016, parmentier2021, roman2017, roman2019, lines2019}, the different conditions prevailing in warm Jupiters necessitate a thorough exploration of the interactions between clouds, radiative transfer, and atmospheric circulation. Particularly, now that JWST enables the characterisation of both transmission and emission spectra of warm gas giants, such an exploration can be guided by the observations. 

Warm gas giants, with equilibrium temperatures of 500–1000 K, are especially interesting because they can host molecules such as water, methane, and carbon dioxide, as well as clouds from various condensates. Studying them is important for understanding the diversity of planetary atmospheres and the processes that shape them. With JWST now providing both transmission and emission spectra, these planets can be explored in greater detail, revealing how their compositions and structures are influenced by their host stars and formation histories.

WASP-80b is one such warm gas giant. It is Jupiter-sized ($\sim$0.97 R$_J$), orbits an M dwarf ($T_{\rm *}$ = 4143 K) on a 3.06 day period, leading to an equilibrium temperature of $\approx 820 K$ \citep{triaud2013}. It has been targeted by JWST as part of the MANATEE JWST Guaranteed Time Observation (GTO) programme (JWST-GTO-1177 \& JWST-GTO-1185). It was observed both because of its favourable observational metrics and to test whether gas giants around M dwarfs have different formation pathways than gas giants around other stars \citep{triaud2023}. This is, to date, the best-characterised warm giant, with both emission and transmission spectra observed over the 2.45 to 11 micron range. 

WASP-80b has already revealed some of its secrets. In \cite{bell2023}, methane was detected in both the dayside and the limb of the planet, making it one of the most robust methane detections to date. Furthermore, 1D retrieval studies of the dayside emission spectra \citep{wiser2025} imply a three times higher solar metallicity and slightly sub-solar C/O ratio, consistent with the current bulk population of hot Jupiters. Arnold et. al., in prep., will analyse and characterise the transmission spectrum in detail. 
In the optical and near-infrared wavelength regime (<2.5 $\mu$m), \cite{jacobs2023} used data from the Wide Field Camera 3 (WFC3) aboard the Hubble Space Telescope to provide constraints on the presence and properties of atmospheric aerosols. \cite{jacobs2023} found that aerosol composition varies across the planet. If clouds are present on the dayside, they may extend into deeper layers of the atmosphere, which results in low geometric albedo (A$_g$ < 0.33). Similar results were obtained by \cite{morel2025}; they also studied the eclipse spectrum of WASP-80b obtained with JWST NIRISS/SOSS (0.68--2.83 $\mu$m). Using retrievals and 1D cloud models, they reject MnS and silicate clouds while concluding that cloud species with weak-to-moderate near-infrared reflectance, along with soots or low-formation-rate tholin hazes, are consistent with the eclipse spectrum. 

Interestingly, whereas most of the dayside spectrum appears cloud-free, grid-based atmospheric retrievals point to a homogeneous, grey cloud to reduce the planetary emission around 4.3 and 10 micron. The 1D grid-based retrieval of \cite{wiser2025} suggests a high internal effective temperature (T$_{\rm int}$) of 381$^{+38}_{-39}$ K and vertical quenching of CH$_4$ with log(K$_{zz}$)=9.13 cm$^2$s$^{-1}$ to explain the observed reduced CH$_4$ abundance, similar to the inference in the warm exoplanet WASP-107b \citep{sing2024, welbanks2024}. However, when clouds are assumed to be completely absent from the atmosphere, the grid-based retrieval favours a lower T$_{\rm int}$ (around 250 K) and a higher metallicity ([M/H]=1.15). As a consequence, the presence or absence of clouds in WASP-80b has direct implications for the measurement of its bulk elemental abundances. 

The presence of clouds should be ubiquitous for planets in the 500 K to 1000 K temperature range. Indeed, within this range, KCl and Na$_2$S clouds are expected to condense (Fig. \ref{fig:ptp_comb}: top left). If the T$_{\rm int}$ is high enough, silicate clouds can also condense in the deep atmospheric layers (Fig. \ref{fig:ptp_comb}: bottom left). However, the observational effect of clouds strongly depends on their spatial distribution, which is determined by the balance between vertical settling and 3D atmospheric mixing. Even though a grey homogeneous cloud opacity can be favoured in 1D retrieval models, it is unclear whether such clouds may indeed exist, given the complex interactions between 3D mixing and settling. 

Clouds are expected to play a crucial role in interpreting our observations of exoplanet atmospheres. First, they affect the energy balance of the atmosphere in a complex way; dayside clouds tend to cool the atmosphere by increasing the albedo, whereas nightside clouds tend to increase the temperature of the atmosphere and reduce the day-to-night heat transport through their greenhouse effect. At the same time, nightside clouds raise the photosphere to higher, colder layers of the atmosphere, which lowers the apparent brightness temperature. Clouds can further bias our inference of atmospheric abundances. As shown by \cite{line2016}, the presence of inhomogeneous clouds at the limb can mimic the effect of a high mean-molecular-weight atmosphere. The presence of inhomogeneous clouds at the terminator of warm giant planets was recently confirmed observationally by \cite{murphy2024}. Finally, clouds, even if they form in regions that are not observable directly (e.g. deeper than the photosphere or on the planetary nightside), can sequester specific elements and affect our determination of bulk elemental abundance ratios, such as C/O. 

General circulation models are necessary to perform a comprehensive analysis of the planet's atmosphere because they can simulate the complex interactions between radiation and dynamics. 
The 3D structure of the planet makes the process of characterising the exoplanet atmosphere from its spectra challenging. The atmospheric properties can vary on different parts of the planets, which is usually averaged in observed spectra. With the transmission and emission available for WASP-80b, a GCM will be a perfect tool to study the global atmosphere.  

For this, we perform 3D general circulation models (GCMs) of WASP-80b with radiatively active cloud tracers with the ADAM framework (previously known as SPARC/MITgcm). Compared to previous studies for these kinds of planets, we can benchmark our modelling framework on the high-quality emission and transmission spectra (Section \ref{subsection: data}). We present our methods in Section \ref{section: methods}, discuss the cloudless case in Section \ref{section: cloudless}, the effect of clouds on the atmospheric dynamics of WASP-80b in Section \ref{section: cloudy}, compare our results to observations by JWST in Section \ref{subsection: spectrum} and, finally, conclude on the possible cloud species present in WASP-80b atmosphere in Section \ref{section: conclusion}.

\section{Methods}
\label{section: methods}

In this section, we discuss the data used in this study and describe the GCM, cloud implementation in the GCM, the post-processing, and the convergence criterion for GCMs. 

\subsection{Data} 
\label{subsection: data}

Both transmission and emission spectra of WASP-80b were observed with the NIRCam and MIRI instruments on board JWST, providing continuous spectra from 2.45 to 14 micron. They consist of NIRCam F322W2 (taken 2022 Oct 29; JWST-GTO-1185 Observation 4), NIRCam F444W (taken 2023 Jun 13; JWST-GTO-1185 Observation 5), and MIRI LRS (taken 2022 Sep 25; JWST-GTO-1177 Observation 2). In this paper, we use the emission spectra published in \cite{wiser2025} and the transmission spectra that will be published in Arnold et. al., in prep. 

These observations led to the detection of the major carbon and oxygen-bearing species expected in an H$_2$-dominated atmosphere at these temperatures. CH$_4$, CO, CO$_2$, H$_2$O, and possibly NH$_3$ are expected in both emission and transmission spectra, with CS$_2$ potentially observed only in transmission.
The presence or absence of clouds could not be constrained with high confidence. From the emission spectrum, the dayside effective temperature was estimated to be $T_{\rm day}=859.9^{+7.1}_{-9.1} K$. The heat redistribution parameter  (from \citealt{wiser2025}), $f=T_{\rm day}/T_{\rm eq}=1.04$ \citep{roth2024} (see Section \ref{section: cloudless} for more details).
\subsection{Abundances}
\label{subsection: Abundances}

Warm Jupiters are less likely to follow thermochemical equilibrium compared to hot Jupiters, primarily due to their lower atmospheric temperatures. At these cooler temperatures, chemical reaction rates are significantly slower, allowing disequilibrium processes such as vertical mixing and photochemistry to dominate the chemical composition \citep{mukherjee2025}. In contrast, the higher temperatures in hot Jupiters allow faster reaction rates, enabling the atmosphere to maintain equilibrium despite dynamical transport. As a result, species such as CH$_4$, CO, NH3, and HCN in warm Jupiter atmospheres often show deviations from equilibrium predictions \citep{drummond_implications_2020, zamyatina2022, zamyatina_quenching-driven_2024, lee2023a}.

\begin{figure}
    \includegraphics[width=\columnwidth]{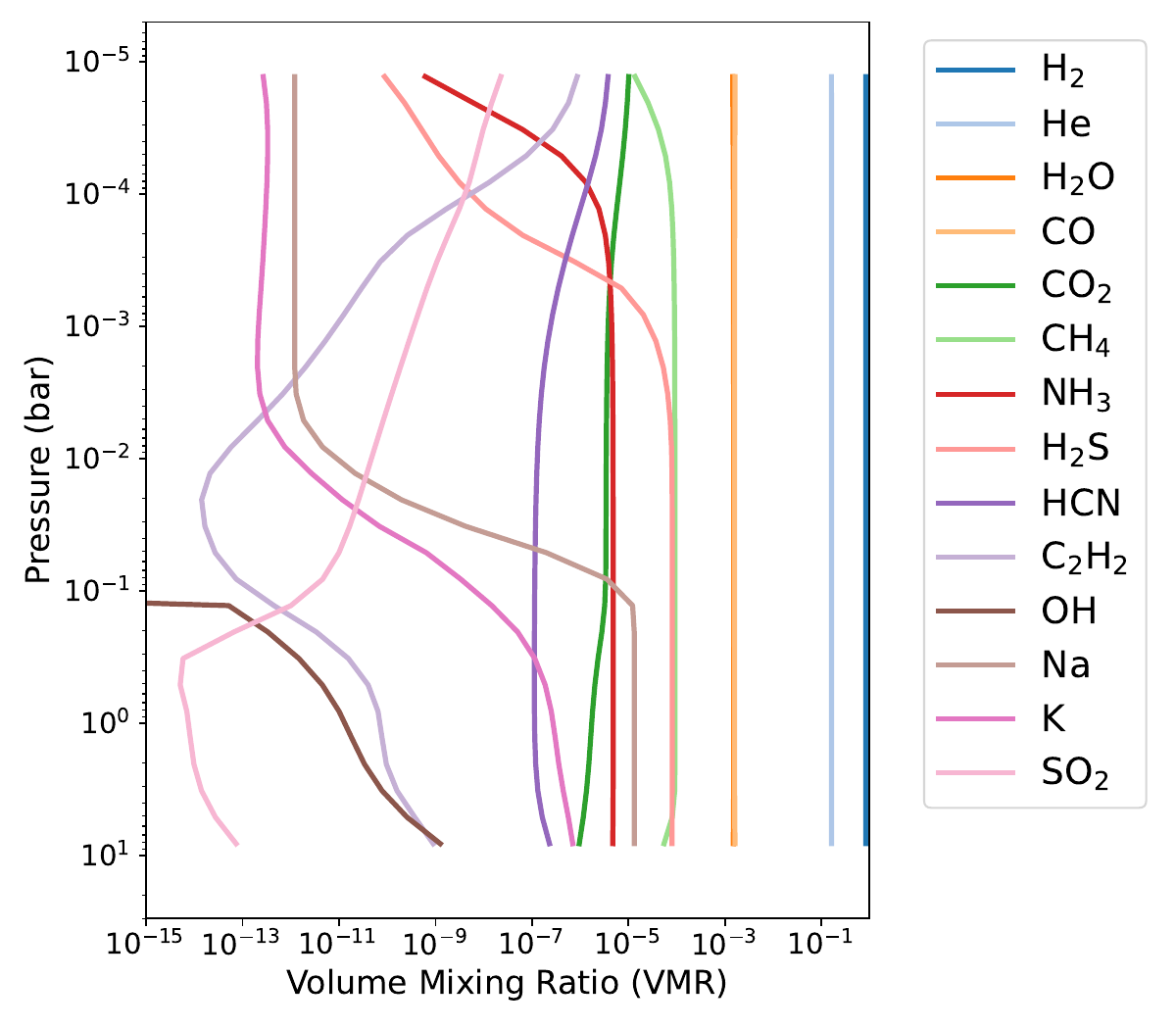}
    \caption{Vertically varying chemical abundance profiles in the atmosphere of WASP-80b, based on the best-fit retrieval from \cite{wiser2025}. The profiles are applied in each GCM column assuming horizontal homogeneity. The retrieved Volume Mixing Ratio (VMR) profiles were interpolated onto the GCM pressure grid and extrapolated beyond the retrieval range using the nearest available values.}
    \label{fig:vmr}
\end{figure}

\cite{wiser2025} and Arnold et. al., in prep., worked on the emission and transmission dataset of WASP-80b from the MANATEE JWST GTO programme. 
In order to reduce the parameter space to be explored with our models, we decided to set the chemical abundances in the model following the atmospheric retrieval work of \cite{wiser2025} with a metallicity of 0.55 and C/O of 0.48. The molecular abundances include H$_2$, He, H$_2$O, CO, CO$_2$, CH$_4$, NH$_3$, H$_2$S, HCN, C$_2$H$_2$, OH, Na, K and SO$_2$, as shown in Fig. \ref{fig:vmr}. The grid retrieval was performed using a grid of ScCHIMERA models (presented in detail in \citealt{iyer2023,wiser2024}), solving for 1D radiative-convective equilibrium that includes photochemistry and disequilibrium chemistry through the VULCAN code \citep{tsai2021}. The best-fit model from \cite{wiser2025} suggests chemical abundances that vary with pressure. We further assume that the horizontal mixing leads to horizontally homogeneous abundances, and thus we apply the 1D vertically varying abundance at every GCM column. Since clouds are implemented as tracers in both the gas phase and condensed phase, their spatial distribution can be inhomogeneous.

\subsection{Opacity}
\label{subsection: opacity}

The original setup of ADAM uses a pre-computed k-coefficient table that assumes equilibrium chemistry from \cite{lupu2021} by calculating the correlated-k coefficients using pre-mixed opacities for each metallicity-C/O combination, as described in \cite{marley2021}.
Given the need to use disequilibrium chemistry abundances, we developed a new module to pre-compute k-coefficient opacity tables for arbitrary chemistry inputs.  

This module calculates the opacity table by mixing the correlated-k tables according to the given abundances, which was then used in the GCM as the atmospheric opacity. 
The correlated-k tables are obtained from \cite{freedman2014,gharib-nezhad2024}. These correlated-k tables are computed on a pressure–temperature (P–T) grid identical to that used by \citet{gharib-nezhad2024} and have already been implemented in \cite{mukherjee2024,mukherjee2025}. \texttt{exo\_k} \citep{leconte2021} is a Python library to handle radiative opacities from various sources for atmospheric applications. We used \texttt{exo\_k} to bin the opacity data into 11 bands, following the approach of \citet{kataria2016}. The resulting binned opacity tables were then combined according to the abundances of the best-fit model from \citet{wiser2025}.

The random overlap method from \cite{lacis1991} was used to mix the opacity tables.
$k$-coefficients are computed individually for each gas, and $k$-coefficients for different gases are combined assuming that the absorption coefficient of one gas $i$, is uncorrelated to that of a second gas $j$, i.e. that their lines are randomly overlapping. The total transmission of the gas mixture over some column density $(u_i,u_j)$ was then given by a simple scalar product:
\begin{equation}
\mathcal T(u_i, u_j) = \mathcal T_i(u_i) \times \mathcal T_j(u_j) .
\label{eq:T_ab}
\end{equation}

Detailed equations and validity of random overlap have been tested in the atmospheres of hot Jupiters and brown dwarfs in \cite{amundsen2017}.
The opacity table mixed by \texttt{exo\_k} was then fixed as the opacity of the atmosphere for the GCM.

\subsection{Internal effective temperatures}

Similar to Jupiter, irradiated giant planets contract and cool; external irradiation experienced by hot or warm Jupiters can slow the cooling but not halt it, due to the development of a radiative zone in the planet’s deep atmosphere \citep{guillot1996}. Atmospheric models must therefore account for both the incoming stellar irradiation and the intrinsic flux due to the planet’s cooling. The latter is measured in terms of an internal effective temperature $T_{\rm int}$, corresponding to a heat flux at thermal wavelengths $\sigma T_{\rm int}^4$. While standard models generally predict a value of $T_{\rm int}$ close to 100 K (the Jupiter value) for planets of Jupiter’s mass and radius, it has been noticed that higher values are required for a number of so-called hot Jupiters \citep{guillot2002}. A number of processes have been invoked to explain the mismatch \citep[e.g.][]{fortney2021, guillot2023}, the most well-known being the interaction between winds in a partially ionised atmosphere and the magnetic field of a tidally locked planet, a process known as ohmic dissipation \citep{batygin2010}. A parametric study to reproduce on average the sizes of transiting hot and warm Jupiters lead to values ranging from $T_{\rm int}$ = 100 K for low-irradiation planets with $T_{\rm eq}$ = 800 K to up to $T_{\rm int}$ = 700 K for hot Jupiters with $T_{\rm eq}$ = 1800 K \citep{thorngren2019}. 

Given WASP-80b’s relatively low $T_{\rm eq}$ value, we would expect a value of $T_{\rm int}$ towards the lower range of these values. We therefore adopt $T_{\rm int}$ = 100 K as our baseline model. However, given the uncertainties in our knowledge of the inflation mechanism and since some atmospheric retrievals point to higher values, we adopt $T_{\rm int}$ = 381 K \citep{wiser2025} as a second possibility.

\subsection{General circulation model}
\label{subsection: gcm}

In this work, we introduce the term ADAM (ADvanced Atmospheric MITgcm) as an umbrella designation for a suite of exoplanet modelling frameworks built upon the MITgcm. This model couples the MITgcm, a GCM maintained at the Massachusetts Institute of Technology \citep{adcroft2004}, and the plane-parallel radiative transfer code of \citep{marley1999} and solves the primitive equations on a cube-sphere grid. Under this naming convention, model configurations are referenced by specifying the relevant physical modules, for example, ADAM with the SPARC module, ADAM with the double-grey radiative module, ADAM with double-grey and H$_2$-dissociation, or ADAM with SPARC, active tracer clouds, and opacity modules, and so on. This nomenclature provides a unified and transparent framework for describing the growing suite of MITgcm-based exoplanet modelling capabilities.
To avoid ambiguity with other MITgcm-based exoplanet modelling efforts, we clarify that ADAM refers specifically to the lineage of models descending from the developments initiated by \cite{showman2009} and extended through a series of key advancements. These include the first non-grey radiative transfer scheme \citep{showman2009}; hydrological-cycle applications to giant planets \citep{lian2010}; extensions to eccentric hot Jupiters \citep{lewis2012PhD}; the 11-bin radiative transfer scheme \citep{kataria2013}; tracer transport on hot Jupiters \citep{parmentier2013a}; Newtonian cooling and double-grey parameter-space explorations \citep{komacek2016, komacek2017}; H$_2$ dissociation physics \citep{tan2019}; active cloud tracers \citep{tan2021, tan2021a, komacek2022a};  haze transport and radiative feedback implementations \citep{steinrueck2023} and the implementations in this work for custom molecular abundances. \footnote{The adoption of the name ADAM honours the late Adam Showman, whose pioneering contributions laid much of the conceptual and practical groundwork for modern exoplanet atmospheric dynamics. The name was first proposed by Maria Steinrueck.}

We use ADAM with SPARC, active tracer clouds, and custom abundance modules to simulate the atmosphere for WASP-80b. Following \cite{mayne2014}, we verified the validity of the four main approximations used to derive the primitive equations. 
This model has been successfully applied to a wide range of exoplanets, including hot Jupiters \citep{showman2009,showman2015,liu2013,parmentier2016,parmentier2018,parmentier2021,kataria2015,kataria2016,lewis2017,steinrueck2019,tan2021}, highly eccentric hot Jupiters \citep{kataria2013,lewis2014}, warm Jupiters \citep{showman2015} and super-Earths \citep{kataria2014,zhang2017}. 

We initialised all the GCMs with the parameters from Table \ref{tab:gcm_par}, unless stated otherwise.
The temperature profile was initialised using the analytical model of \cite{parmentier2015} that uses the analytical expression of \cite{parmentier2014} adjusted to represent the global average temperature profile of solar-composition atmospheres \citep{fortney2007}. The simulations did not incorporate any explicit Rayleigh drag-in, as discussed in \cite{showman2009}. The temperature and velocity fields are subjected to a fourth-order horizontal Shapiro filter, which performs the primary damping \citep{parmentier2021}. The atmospheric circulation that is fuelled by huge, planetary-scale wave flow interactions should not be greatly impacted by the Shapiro filter since it dissipates kinetic energy by smoothing horizontal gradients at the grid level \citep{showman2011, hammond2018}. 

\begin{table}[ht]
\centering
\small
\caption{GCM parameters used in this study.}
\begin{tabular}{l c}
\hline
\hline
\multicolumn{2}{c}{GCM parameters} \\
\hline
Reference surface pressure [bar]       & 200        \\
Pressure range                         & 200 bar -- 2 $\mu$bar \\
Upper boundary pressure [bar]          & $2 \times 10^{-6}$  \\
Specific heat capacity $c_p$ [J\,kg$^{-1}$\,K$^{-1}$] & $1.3 \times 10^{4}$  \\
Adiabatic coefficient $\kappa$         & 0.286      \\
Gravity $g$ [m\,s$^{-2}$]              & 13.96      \\
Semi-major axis [AU]                   & 0.034      \\
Planetary radius $R_p$ [m]             & $6.79 \times 10^{7}$ \\
Eccentricity $e$                       & 0          \\
Orbital period [days]                  & 3.0678     \\
$\log_{10}$ (metallicity)              & 0.55       \\
Hydrodynamic timestep [s]              & 40         \\
\hline
\end{tabular}

\tablefoot{
We use a cubed-sphere resolution of C32 (128\,$\times$\,64 in longitude\,$\times$\,latitude).  
The model includes 53 pressure levels, providing nearly three levels of resolution per scale height.
}

\label{tab:gcm_par}
\end{table}

\subsection{Clouds}
\label{subsection: clouds}
Clouds are included in the GCMs in the form of tracers. These tracer-based clouds are radiatively active and evolve according to the local temperature and pressure. The tracers move in the atmosphere according to the atmospheric circulation. The tracers are then given the properties of the specific cloud species, such as condensation curve, density, molecular weight, refractive indices, and Mie parameters \citep{natasha_batalha_2020_5179187}. All the cloud species are initialised with the solar abundance \citep{lodders2003} scaled to the metallicity of the planet. The radiative feedback of clouds is implemented in the model by computing their optical properties (optical depth, single scattering albedo, and asymmetry factor) from Mie efficiency data \citep{natasha_batalha_2020_5179187} (Na$_2$S: \citealt{Montaner1979, Khachai2009}; KCl: \citealt{Querry1987OpticalCO};  MgSiO$_3$: \citealt{Scott1996}). These properties are then included in the radiative transfer scheme, allowing clouds to interact with radiation and modify the heating rates. The cloud properties are given in Table \ref{tab:cloud-properties}. Fig. \ref{fig:ssa} presents the cloud single scattering albedo, extinction, and absorption opacities normalised per unit mass. 

We couple the primitive equations of motion to tracer equations for the transport of condensible vapour and cloud condensate by following the method of \cite{tan2021,tan2021a}:
\begin{equation}
\label{eq:vapour}
    \frac{dq_v}{dt} = (1 - \delta)\frac{\mathrm{min}(q_s - q_v, q_c)}{\tau_\mathrm{c}} - \delta\frac{\left(q_v - q_s\right)}{\tau_\mathrm{c}} - \frac{q_v - q_\mathrm{deep}}{\tau_\mathrm{deep}} \mathrm{,} 
\end{equation}
\begin{equation}
\label{eq:condensate}
    \frac{dq_c}{dt} = \delta \frac{\left(q_v-q_s\right)}{\tau_\mathrm{c}} - (1-\delta) \frac{\mathrm{min}(q_s - q_v, q_c)}{\tau_\mathrm{c}} - \frac{\partial\left(q_c \langle V_\mathrm{p}\rangle \right)}{\partial p} \mathrm{}
\end{equation}
where $q_v$ is the mass mixing ratio of condensible vapour relative to the background air, $q_c$ is the mass mixing ratio of cloud condensate particles, $q_s$ is the mass mixing ratio of condensible vapour at saturation, $\delta$ is the supersaturation indicator, which is set to one when vapour is super-saturated and zero if vapour is sub-saturated, $\tau_\mathrm{c}$ is the cloud and condensible vapour tracer relaxation timescale, the deep vapour mass mixing ratio $q_\mathrm{deep}$, and the deep vapour replenishment timescale $\tau_\mathrm{deep}$. The particle size of the clouds has a log-normal size distribution. The terminal settling velocity in pressure coordinates that is averaged over the particle size distribution $\langle V_\mathrm{p}\rangle$. The mean particle size is given as an input to the GCM. For a given particle size, the cloud settling velocity $V_s$ as a function of pressure and temperature was calculated using  Eqs. (3-7) of \cite{parmentier2013a}.

Adding clouds to ADAM was done in the same way as \cite{tan2021,tan2021a}. The cloud module of ADAM has already been applied in \cite{komacek2022} for a two-stream double-grey radiative transfer scheme, whereas we apply it using a non-grey radiative transfer. 

\begin{figure*}
    \centering
    \includegraphics[width=\textwidth]{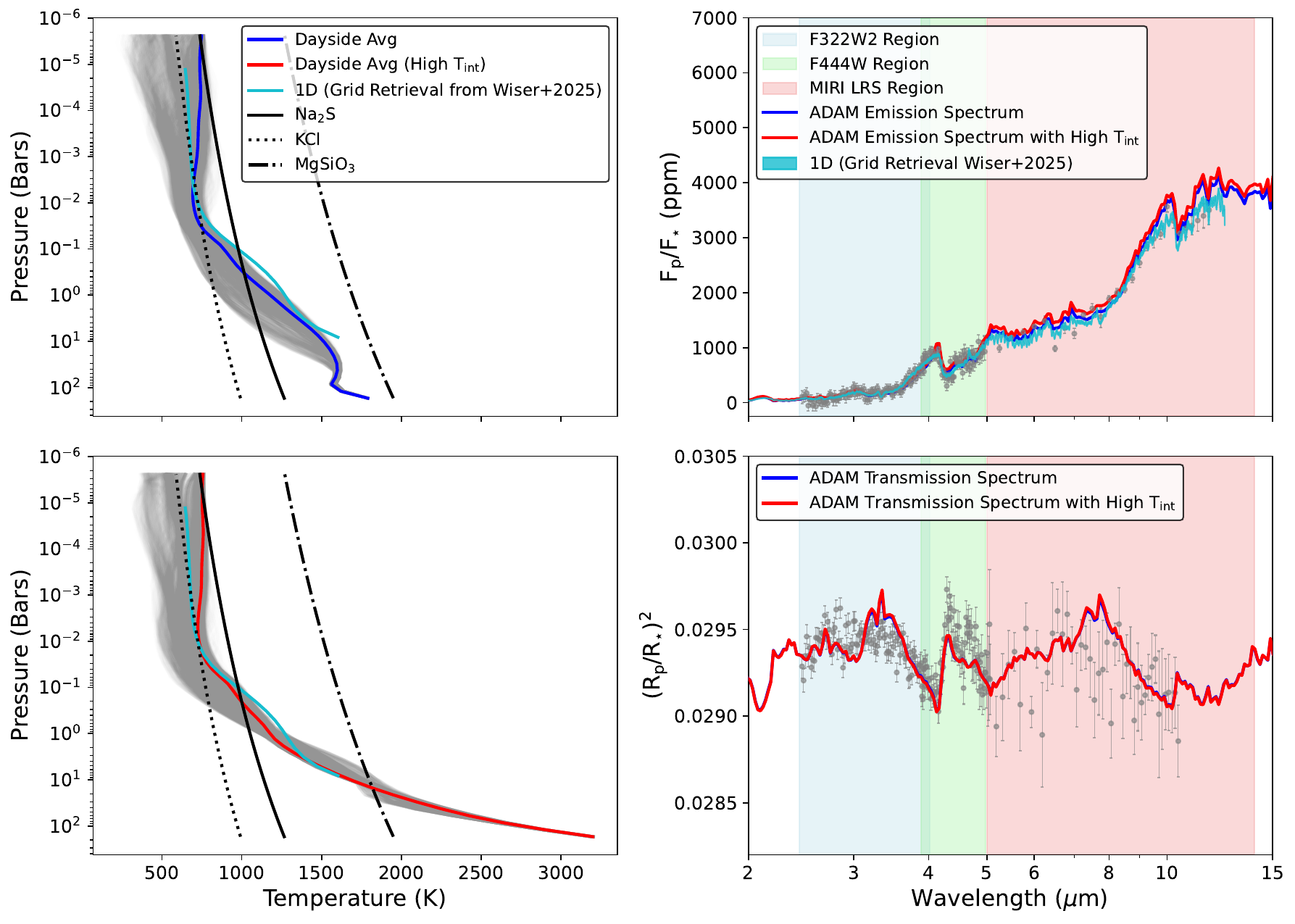}
    \caption{Left Column: Pressure-temperature profiles from the cloudless GCM models (top, cloudless GCM with low T$_{\rm int}$; bottom, cloudless GCM with high T$_{\rm int}$; grey, profiles at all latitudes and longitudes and light blue, 1D profile from \citealt{wiser2025}). The condensation curves of several important species are plotted as dashed lines. Right Column: ADAM spectrum plotted with JWST observations in grey (top, emission spectra for cloudless GCM with low and high T$_{\rm int}$ showing significant overlap, indicating minimal sensitivity to internal heating in the terminator region, plotted along with 1D spectrum from \cite{wiser2025}; bottom, transmission spectra for cloudless GCM with low and high T$_{\rm int}$ showing significant overlap, indicating minimal sensitivity to internal heating in the terminator region).}
    \label{fig:ptp_comb}
\end{figure*}

As seen in Fig. \ref{fig:ptp_comb}, the condensation curves of Na$_2$S and KCl cross the pressure-temperature profile. Hence, these clouds are included. And following the results of \cite{wiser2025}, we include MgSiO$_3$ clouds as discussed in the introduction. 
The condensation curve of Na$_2$S (Eq. 28 from \citealt{visscher2006}), KCl (Eq. 20 from \citealt{morley2012}) and MgSiO$_3$ (Eq. 20 from \citealt{visscher2010}) are given by:
\begin{equation}
    10^4/T_{\rm cond}(\rm Na_2S) \approx 10.05 - 0.72 \log_{10} P_T - 1.08[M/H] 
    \label{eq.Na$_2$S}
\end{equation}

\begin{equation}
    10^4/T_{\rm cond}(\rm KCl) \approx 12.479 - 0.879  \log_{10} P_T - 0.879 [M/H] 
    \label{eq.kcl}
\end{equation}

\begin{equation}
    10^4/T_{\rm cond}(\rm MgSiO_3) \approx 6.26 - 0.35  \log_{10} P_T - 0.7 [M/H] 
    \label{eq.mgsio3}
\end{equation}

\noindent where $T_{\rm cond}$ is equilibrium condensation temperature, $P_T$ is total gas pressure, and $[M/H]$ is the metallicity of the atmosphere. 

We calculated the total gas pressure from the condensation curves. When the local conditions are cooler than the condensation temperature of the cloud species, the gas condenses into clouds. Due to this temperature dependence, clouds are usually expected in the cooler regions of the planet. Instead of calculating the partial pressure of each condensation species tracer and comparing it to the saturation vapour pressure, we directly compare the mass mixing ratio of condensate to the local saturation mass mixing ratio, following \cite{tan2019a}, 
\begin{equation}
    q_s = P_T \times q_\mathrm{deep} / p
    \label{eq.qs}
\end{equation}
\noindent where $P_T$ is the saturation vapour pressure determined by the local temperature from the condensation curve, and $p$ is the local gas pressure in the model atmosphere. 

When the condensible vapour mixing ratio is higher than the required saturation mixing ratio, $q_s$, a cloud forms; otherwise, when the vapour mixing ratio is less than $q_s$, evaporation takes place.
The condensation point depends on the local condensing vapour. The saturation mixing ratio is assumed to be a function of pressure and temperature, so $q_s$ is $q_{\rm deep}$ at the condensation pressure ($p_{\rm deep}$) with a steep decrease when pressure is less than $p_{\rm deep}$. At pressures larger than $p_{\rm deep}$, $q_s$ is assumed to be arbitrarily large such that no condensation would occur.
The vapour field is relaxed towards $q_{\rm deep}$ with the time-scales shown in Table \ref{tab:cloud-properties}. The parameter choices for MgSiO$_3$ differ from those used for the other cloud species due to numerical stability constraints in the model (see Section \ref{section: cldsontemp} for details). Variations in the assumed timescales do not significantly influence the behaviour of the cloud distribution or the resulting observables.

The deep layers of our models reach the unstable convective region. The effects of rapid convective mixing, using a simple convective adjustment scheme as in the NCAR Community Atmosphere Model (\cite{collins2004}, see their Section 4.6), were parametrised. In a vertical atmospheric column, any two adjacent layers that are unstable are instantly converted to a neutral convective condition while still conserving the total sensible heat $\sum \Delta p T$, where $\Delta p$ is the pressure-dependent layer thickness.  The entire column is periodically scanned in a single dynamical step until convective instability is removed everywhere.  During adjustment, tracers are also uniformly distributed over the modified domain. The horizontal direction is not adjustable. The GCM models do not include an upper sponge layer, which is typically used to minimise the effects of wave reflection from the upper boundary on dynamics at levels of interest. The physical phenomenon seen in our results and conclusions of this study should be independent of the numerical setup of the numerical diffusion, as shown in \cite{tan2021}.

\begin{table*}[ht]
\centering
\setlength{\tabcolsep}{12.5pt}
\caption{Input parameters for cloud microphysics in the GCM simulations.}
\begin{tabular}{llll}
\hline
\hline
\multicolumn{4}{c}{Cloud properties} \\ \hline
 & Na$_2$S & KCl & MgSiO$_3$ \\ \hline

Condensate vapour deep mixing ratio ($q_{\rm {deep}}$) 
& $5.9 \times 10^{-5} \times 10^{[Z]}$ 
& $7.13 \times 10^{-6} \times 10^{[Z]}$ 
& $0.0027 \times 10^{[Z]}$ \\

Condensate vapour source pressure ($p_{\rm {deep}}$) 
& $10^7$ 
& $10^7$ 
& $3.3 \times 10^7$ \\

Condensate vapour relaxation timescale ($\tau_{\rm {c}}$) 
& 100 s 
& 100 s 
& 200 s \\

Condensate vapour deep relaxation timescale ($\tau_{\rm {deep}}$) 
& 1000 s 
& 1000 s 
& 100 s \\

Condensate density ($\rho_c$) \citep{roman2021}
& 1860 kg/m$^3$ 
& 1980 kg/m$^3$ 
& 3190 kg/m$^3$ \\

Mean particle size ($r_0$) 
& [0.1, 1, 5, 10] $\mu$m 
& [0.1, 1, 5, 10] $\mu$m 
& [0.1, 1, 5, 10] $\mu$m \\

Log-normal distribution width ($\sigma$) 
& 1.65 
& 1.65 
& 1.65 \\

Internal effective temperature T$_{\rm int}$ (K) 
& 100 
& 100 
& 381 \\ \hline
\end{tabular}

\tablefoot{
Metallicity is represented by [Z], with a value of 0.55 for WASP-80b \citep{natasha_batalha_2020_5179187}.  
The listed T$_{\rm int}$ values correspond to those used in the initialisation of the GCM model.  
Optical constants are taken from: Na$_2$S -- \citet{Montaner1979, Khachai2009},  
KCl -- \citet{Querry1987OpticalCO},  
MgSiO$_3$ -- \citet{Scott1996}.
}

\label{tab:cloud-properties}
\end{table*}

\subsection{Post-processing}
\label{subsection: post-processing}

The post-processing of the GCMs was done using the \texttt{gcm\_toolkit} code. \texttt{gcm\_toolkit} is an open-source Python package to read, post-process, and plot 3D GCM data. This package has already been implemented for GCM studies for exoplanets in \cite{carone2020,schneider2022}. The \texttt{gcm\_toolkit} regrids the output to a regular latitude-longitude grid with the same number of layers: 53. The new latitude-longitude grid is regridded to (45, 72). The output of \texttt{gcm\_toolkit} was then used as an input to the 3D radiative transfer code \texttt{gcm\_toolkit}. 

\texttt{gCMCRT} is a 3D GPU-accelerated Monte Carlo Radiative Transfer \citep{lee2022} that is used on a spherical geometry grid to post-process the GCM results. \texttt{gCMCRT} has been previously used along with ADAM at low resolution \citep{komacek2022} and high resolution \citep{wardenier2021}. The effects of cloud structures on the resulting transmission spectra and emission spectra of the model output are studied using \texttt{gCMCRT}. \texttt{gCMCRT} simulates the path of photon packets through a planetary atmosphere in 3D. It is a hybrid MCRT and ray-tracing code, using the 'peel-off' ray-tracing method to produce images and spectra of the simulated planet. 

Previous post-processing of GCMs was usually done by using a plane-parallel code and produced a planetary flux per unit surface, which is then scaled by the measured planetary radius in transit. On the contrary, \texttt{gCMCRT} outputs the total outgoing flux, and for gas giant planets, there is no satisfactory definition of surface to normalise the flux. However, an error in the planetary radius would directly translate into an error in the outgoing flux; we therefore need to make sure our model has the correct radius.
It is important to note that the radius can be used as a free parameter after the spectrum is obtained. But in this case, the radius of the planet and corresponding reference pressure are calculated by fitting the transmission spectrum of the planet to the observations. 

The GCM usually takes as input the radius at 200 bar, which we take as the observed planetary radius. Although this is inaccurate, the atmospheric dynamics are not strongly dependent on small changes in radius. However, when computing the emission spectra, the error on the planetary size is larger than the JWST error bars. As a consequence, for the emission spectra calculation, we need to benchmark the planetary radius. For this, we calculate a transit spectrum with \texttt{gCMCRT} with our initial 200 bar radius (0.97 R$_J$). We then determine the radius difference between the observed and modelled radius and adjust our 200 bar radius accordingly, so that the calculated transit spectra match the observations. This results in a radius of 0.92 R$_J$ is the 200 bar radius that is then used to calculate the emission spectra.
This radius is fixed for the rest of the study, unless otherwise stated. The use of an accurate radius is important as no 'Photospheric Radius Correction factor' $(R_p/R_s)^2$ \citep{fortney2019} was used. \texttt{gCMCRT} avoids this factor by directly accounting for the volume of the emitting regions. \texttt{gCMCRT} correctly weighs the contribution of different temperatures to the spectrum, avoiding known biases \citep{feng2016,taylor2021}.

We also use \texttt{gCMCRT} to post-process the GCMs with clouds. We parametrise a log-normal cloud particle size distribution in our GCM as
\begin{equation}
\label{eq:cldparsize}
    \frac{dn_c}{dr} = \frac{n_\mathrm{c}}{\sqrt{2\pi}\sigma r} e^{-\left[\mathrm{ln}(r/r_0)\right]^2/(2\mathrm{ln}(\sigma)^2)} \mathrm{,}
\end{equation}
where $n_c$ is the cloud particle number per dry air mass, $r_0$ is the mean particle size, $\sigma$ is the log-normal distribution width, and $n_\mathrm{c}$ is the number of cloud particles per dry air mass, calculated as
\begin{equation}
    n_\mathrm{c} = \frac{3 \rho q_c}{4\pi \rho_c r_0^3 e^{(4.5(ln(\sigma))^2)}}
\end{equation}
where $\rho$ is the density of the atmosphere. Our assumed cloud particle size distribution affects the vertical settling of particles through the sink term for cloud condensate on the right hand side of Eq.~\eqref{eq:condensate}, $- \frac{\partial\left(q_c \langle V_\mathrm{p}\rangle\right)}{\partial p}$, in which the mean terminal velocity is obtained from proper averaging over the size distribution in Eq.~\eqref{eq:cldparsize}. 

\subsection{Convergence test}
\label{subsection: congvergence}

\begin{figure}[t]
    \centering
    \includegraphics[width=\columnwidth]{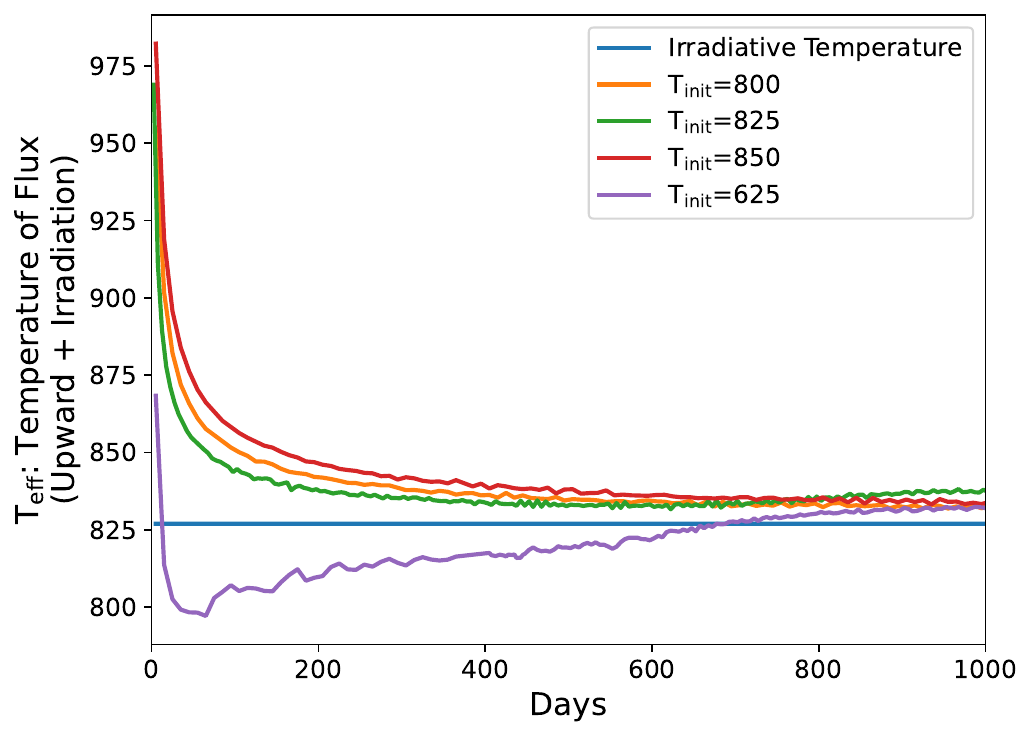}
    \caption{Effective temperature calculated from global outgoing flux as a function of simulation time for different initial temperature-pressure profiles ($T_{\rm eq}$ = 625, 800, 825, and 850 K). The convergence of all cases towards the same equilibrium flux demonstrates that the GCM solution is independent of the initial conditions.}
    \label{fig:pconvergence}
\end{figure}

The duration of each simulation was 1000 days. A fully converged state would require thousands of days for the models to be integrated \citep{mendonca2018,mendonca2020,wang2020}. This has several consequences.  First, because the integration period is too short, the results cannot be affected by the deep boundary condition we chose. The second is that there is insufficient integration time for a deep circulation to emerge and significantly affect the photospheric level \citep{mayne2017,sainsbury-martinez2019,carone2020,wang2020}. However, at photospheric levels, a pseudo-steady state is achieved \citep{showman2009}. Assuming that the deep flow does not substantially alter the equilibrium of photospheric levels, this steady state is accurate. To test the convergence of the model, we use the net flux at the top of the atmosphere (TOA) as a convergence criterion. 
The outgoing flux must be equal to the incoming flux.  
The model is assumed to have reached a pseudo-steady state when the net flux converges to the equilibrium temperature of the planet. For this test, we use different initial temperature-pressure profiles ($T_{\rm eq}$ = 625 K, 800 K, 825 K, 850 K) to show that the GCMs converge to the equilibrium temperature of the planet irrespective of the initial conditions. The convergence test is shown in Fig. \ref{fig:pconvergence}. The test shows that the model converges to the equilibrium temperature of the planet in less than 1000 days. Fig. \ref{fig:pconvergence} shows that there is only a small flux difference after 1000 days. 

For the cloud cases with larger particle sizes (5 and 10 $\mu$m), the sedimentation timescales remain well within the duration of our GCM integrations across the entire modelled pressure range. These clouds are therefore expected to be fully equilibrated with the background mixing. For the smaller particle sizes (0.1 and 1 $\mu$m), the sedimentation timescales are sufficiently short at low pressures that sedimentation, mixing equilibrium is achieved in the photospheric regions: $< 0.1$ bar for 0.1 $\mu$m particles and $< 1$ bar for 1 $\mu$m particles. The vertical profiles for these small grains can therefore be interpreted as dynamically mixed and sedimentation-equilibrated in these upper atmospheric layers. However, at higher pressures, the sedimentation timescale reaches $10^5$, $10^7$ hours, orders of magnitude longer than the duration of our GCM runs, and the deep atmosphere is not expected to have reached a steady-state balance between settling and vertical transport.

\section{Cloudless GCMs}
\label{section: cloudless}

We use ADAM to model the cloudless atmosphere of WASP-80b, using molecular abundances from \cite{wiser2025}. These abundances were derived from 1D radiative-convective-photochemical equilibrium (RCPE) models and provide a chemically consistent baseline for the 3D GCM.

Fig. \ref{fig:ptp_comb} compares the thermal structure of WASP-80b in two cloudless scenarios: one with a low T$_{\rm int}$ (100 K) and another with a high T$_{\rm int}$ (381 K). In the low T$_{\rm int}$ case, the dayside-nightside temperature contrast at the top of the atmosphere reaches about 250 K, but this contrast diminishes rapidly with depth and drops below 50 K around 0.05 bar. Below that, the atmosphere becomes nearly horizontally uniform in temperature, suggesting efficient heat redistribution.

In the cloudless high T$_{\rm int}$ case, the overall thermal structure remains broadly similar, but the dayside temperatures in the upper atmosphere are slightly warmer. The extra internal heat has only a minor effect on the global thermal distribution, and the day-night contrast remains the same. The terminator regions also show temperature asymmetries of less than 150 K in both cases, indicating longitudinal variation at the limb \citep{fu2025}.

\begin{figure*}
    \centering
    \includegraphics[width=\textwidth]{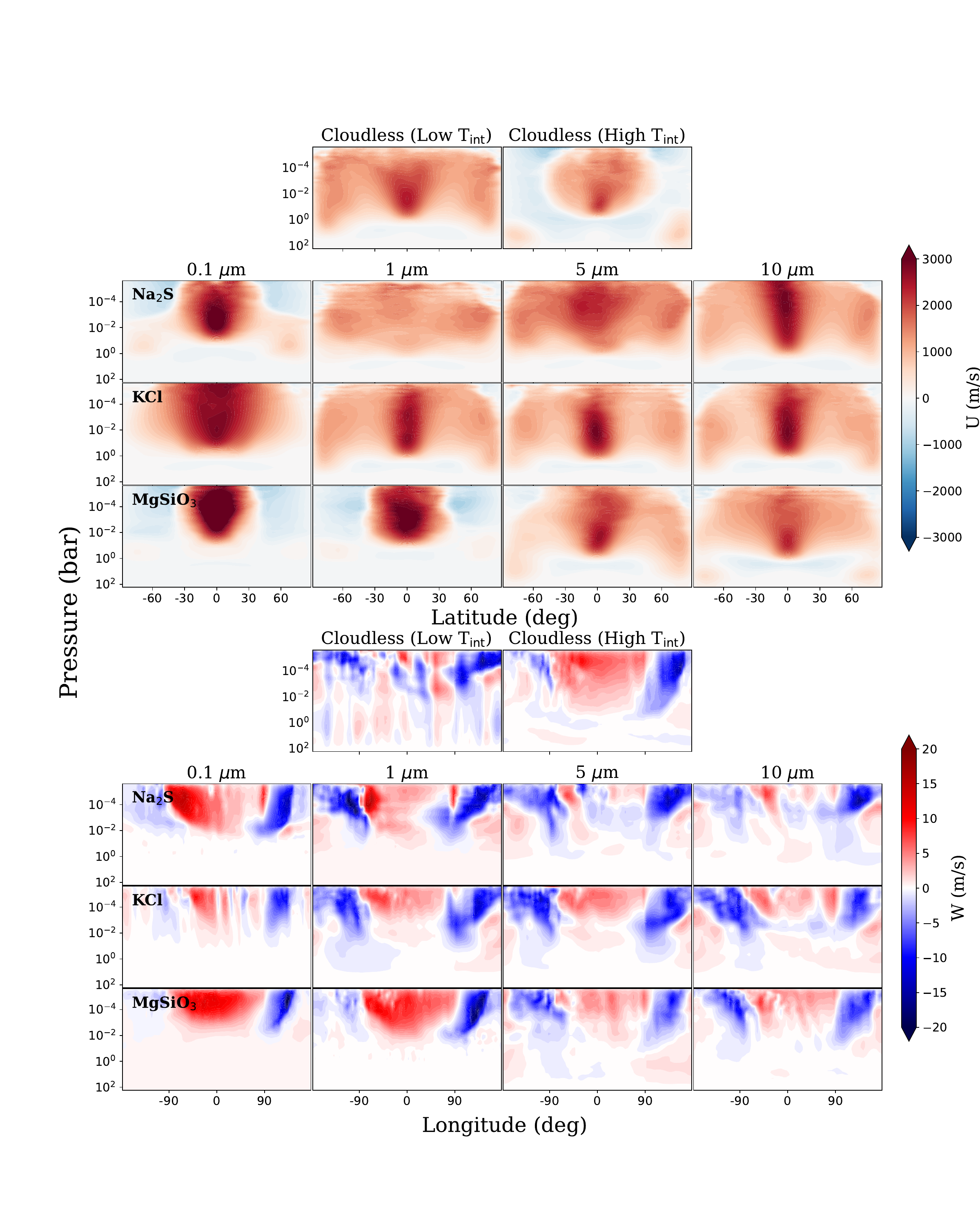}
    \caption{Top: Zonal-mean zonal wind speed (latitudinal distribution of zonal component of wind (U) averaged over longitudes) for all models. Within each subplot, the x-axis shows the latitude and the y-axis the pressure. The first row shows cloudless models, low (100 K) and high (381 K) T$_{\rm int}$. Rows 2, 3, and 4 correspond to Na$_2$S, KCl, and MgSiO$_3$ clouds (with high T$_{\rm int}$), respectively, with each column representing a different particle size indicated at the top. Bottom: Same as top, but for longitudinal distribution of the vertical component of wind (W) averaged over latitudes. The vertical component shows the strength of upwelling on the dayside and downwelling on the nightside for different models.}
    \label{fig:puwmean}
\end{figure*}

The differences in atmospheric dynamics become more apparent when comparing the zonal wind structures (Fig. \ref{fig:puwmean}). In the low T$_{\rm int}$ case, the winds are uniformly prograde across all latitudes. The circulation is dominated by a strong equatorial super-rotating jet and weaker high-latitude jets. This wind pattern is consistent with previous studies of tidally locked hot Jupiters, where thermal gradients drive equatorial jets through eddy momentum transport \citep{showman2011}.

In contrast, the high T$_{\rm int}$ case exhibits a more complex wind structure. While the equatorial super-rotating jet still forms, the high-latitude winds reverse direction and become retrograde, breaking the global prograde symmetry seen in the lower T$_{\rm int}$ case. As shown in Fig. \ref{fig:puwmean}, the high T$_{\rm int}$ case develops a pronounced upwelling region on the dayside.

\begin{table*}[t]
\centering
\caption{$\chi^2_{\rm red}$ values for transmission and emission spectra across cloud species and particle sizes.}
\begin{tabular}{ccccccc}
\hline
\hline
 & \multicolumn{2}{c}{Na$_2$S} & \multicolumn{2}{c}{KCl} & \multicolumn{2}{c}{MgSiO$_3$} \\
\hline
 & Transmission & Emission & Transmission & Emission & Transmission & Emission \\
\hline
0.1 & \cellcolor[HTML]{34FF34}2.22 & \cellcolor[HTML]{FD6864}12.52 & \cellcolor[HTML]{34FF34}2.52 & \cellcolor[HTML]{FD6864}13.90 & \cellcolor[HTML]{F8A102}2.86 & \cellcolor[HTML]{FD6864}29.53 \\
1   & \cellcolor[HTML]{F8A102}2.25 & \cellcolor[HTML]{FD6864}22.87 & \cellcolor[HTML]{34FF34}2.06 & \cellcolor[HTML]{34FF34}2.12 & \cellcolor[HTML]{F8A102}2.45 & \cellcolor[HTML]{FD6864}20.03 \\
5   & \cellcolor[HTML]{34FF34}2.43 & \cellcolor[HTML]{FD6864}8.72  & \cellcolor[HTML]{34FF34}2.61 & \cellcolor[HTML]{34FF34}2.13 & \cellcolor[HTML]{34FF34}2.87 & \cellcolor[HTML]{34FF34}1.90 \\
10  & \cellcolor[HTML]{34FF34}2.83 & \cellcolor[HTML]{34FF34}1.87                          & \cellcolor[HTML]{34FF34}2.81 & \cellcolor[HTML]{34FF34}2.14 & \cellcolor[HTML]{34FF34}2.81 & \cellcolor[HTML]{34FF34}2.55 \\
\hline
\hline
 & \multicolumn{2}{c}{Cloudless with low T$_{\rm int}$} & \multicolumn{2}{c}{} & \multicolumn{2}{c}{Cloudless with high T$_{\rm int}$} \\
\hline
 & \cellcolor[HTML]{34FF34}2.75 & \cellcolor[HTML]{34FF34}1.97  &  &  & \cellcolor[HTML]{34FF34}2.90 & \cellcolor[HTML]{34FF34}2.73  \\
\hline
\end{tabular}

\tablefoot{
The reduced chi-square values ($\chi^2_{\rm red}$) quantify the agreement between each model and the JWST transmission and emission spectra.
Models rejected based on $\chi^2_{\rm red}$ are shown in red, models rejected by our feature-based comparison (see Section \ref{subsection: spectrum} for details) are shown in orange, and the models that can predict the observations well are in green.
}

\label{tab:chi-sq}
\end{table*}

As shown in Fig. \ref{fig:ptp_comb}, the cloudless GCM matches very well the emission spectrum of the planet, with a $\chi^2_{\rm red}$ value of 1.97. This is remarkable for a model as complex as a general circulation model without any fine-tuning of the parameters. It validates both the modelling framework and our approach to use the retrieved chemical abundances from 1D retrieval as input for the 3D model. In terms of dynamics, the agreement of the overall level of dayside flux means that the model properly captures the heat transport in WASP-80b, with a heat redistribution parameter (calculated from cloudless GCM case) of $f$ = 1.04 ($f = (T_{\rm day}/T_{\rm eq})^4$) corresponding to an efficient heat transport \citep{roth2024}. $T_{\rm day}$ represents the temperature of a blackbody that emits the same total energy as the planet's dayside. Based on this definition, the redistribution factor (commonly referred to as the $f$-factor) ranges from a minimum of $f$ = 1, which corresponds to complete heat redistribution across the entire planet, resulting in uniform temperatures between the day and night sides to a maximum of $f$ = 2.66, which represents no heat redistribution, with all incoming energy emitted from the dayside alone. An intermediate value, $f$ = 2, represents a theoretical case where heat is redistributed only across the dayside but not to the nightside.

In Fig. \ref{fig:ptp_comb}, we compare the spectrum and thermal profiles of the 1D grid-retrieval of \cite{wiser2025} with our cloudless 3D GCM output. We observe some differences in the temperature-pressure (TP) profiles when comparing the 1D-RCPE (ScCHIMERA) with the 3D-GCM (ADAM). These discrepancies highlight the importance of multidimensional effects, such as atmospheric dynamics and horizontal heat transport, which are inherently captured in 3D models but absent in 1D frameworks. The transport-dominated region ( > 10$^{-1}$ bar) in the 3D model shows cooler temperatures due to efficient horizontal heat redistribution, while the photospheric layers (around 10$^{-2}$ bar) exhibit good agreement between the two models. At lower pressures (< 10$^{-1}$ bar), the 3D model tends to be warmer, maintaining radiative balance with deeper layers.

The major differences that are observed between the spectra are around 4 and 10 $\mu$m, when compared to the 1D spectrum from \cite{wiser2025}. The 4 $\mu$m peak is overestimated by the GCM, whereas in 1D, a grey opacity has been included to obtain a better fit to the data. This grey opacity can be attributed to clouds or molecules such as PH$_3$. If the grey opacity is considered to be from the clouds, the inferred T$_{\rm int}$ becomes highly sensitive to their presence and properties, as discussed in the introduction. In the cloudless GCM case, deeper layers are visible, making high internal effective temperatures incompatible with the observed 4-4.5 $\mu$m flux. The differences around 10 $\mu$m are due to the higher temperature from the GCM in the upper atmosphere. The corresponding transmission spectrum slightly overestimates the CH$_4$ abundance, which can be seen at 3.3 and 7.8 $\mu$m features. This overestimation can be due to small changes in the CH$_4$ abundance between dayside and limbs. This discrepancy between emission and transmission spectra is also observed in \citealt{acuna-aguirre2025}. One possible explanation is the photochemical removal of CH$_4$ from the upper atmosphere; additionally, transport-induced quenching driven by vertical and horizontal mixing may also play a role in setting the CH$_4$ abundance. A comprehensive analysis of the abundance differences inferred from emission versus transmission spectroscopy, together with an assessment of stellar contamination effects arising from star spots in transmission spectra, will be presented in Arnold et. al. (in prep.). 
It should be noted that no differences are observed in the transmission spectra of low and high T$_{\rm int}$; this is because chemistry is not included in the GCM.
The cloudless GCM model is also a very good fit to the transit spectrum with a $\chi^2_{\rm red}$ of 2.75. If we neglect the 4.6 $\mu$m feature, which is likely due to photochemistry, the $\chi^2_{\rm red}$ goes down to 2.67. 

\section{Cloudy GCMs}
\label{section: cloudy}

To understand the role of clouds in shaping the atmosphere, we implement clouds in the GCM as radiatively active tracers that interact with both the thermal structure and the winds. Clouds, which are expected to form in the relatively cool environment of this 820 K warm Jupiter, play a dual role: they are influenced by the large-scale circulation and, in turn, significantly impact the planet's thermal profile and wind dynamics. This radiative feedback is particularly important; it can alter the spectra. In the following sections, we explore how the inclusion of clouds modifies the atmospheric structure and circulation patterns, providing physical insight into the observed JWST spectra of WASP-80b. We carried out separate simulations for each cloud species and mean particle size to determine how they individually affect atmospheric dynamics and spectral features. 

\subsection{Spatial cloud distribution}
\label{subsection: spatial cloud distribution}

\begin{figure*}[!hbt]
    \centering
    \includegraphics[width=\textwidth]{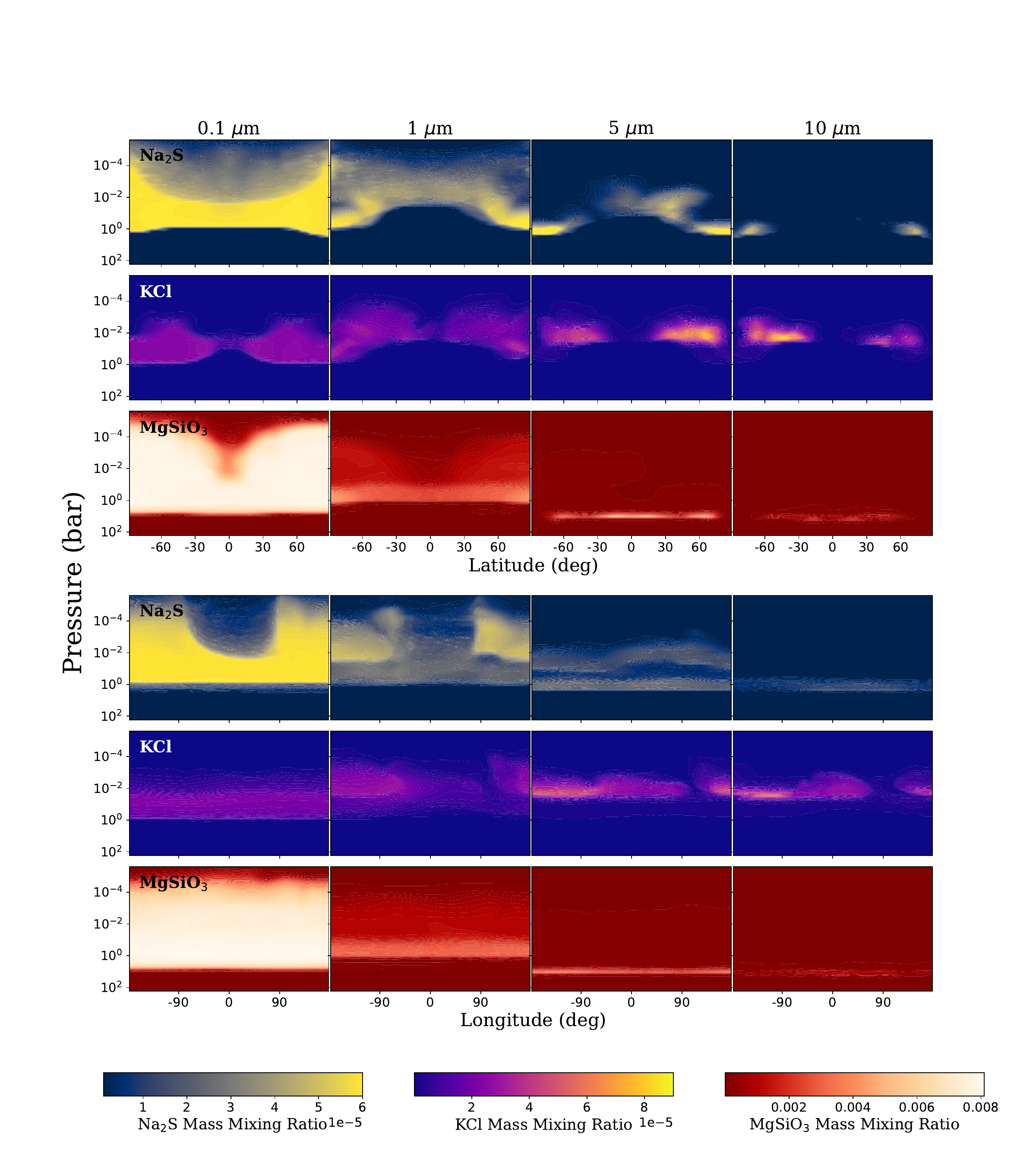}
    \caption{Top: Latitudinal distribution of clouds (shows equatorial and polar regions), averaged over longitudes for all models.  Within each subplot, the x-axis shows the latitude and the y-axis the pressure. Rows 1, 2, and 3 correspond to Na$_2$S, KCl, and MgSiO$_3$ (with high T$_{\rm int}$) clouds, respectively, with each column representing a different particle size indicated at the top. Bottom: Same as top, but for longitudinal distribution of clouds (shows dayside and nightside of the planet), averaged over latitudes for all models.}
    \label{fig:pclds}
\end{figure*}

\begin{figure*}
    \centering
    \includegraphics[width=\textwidth]{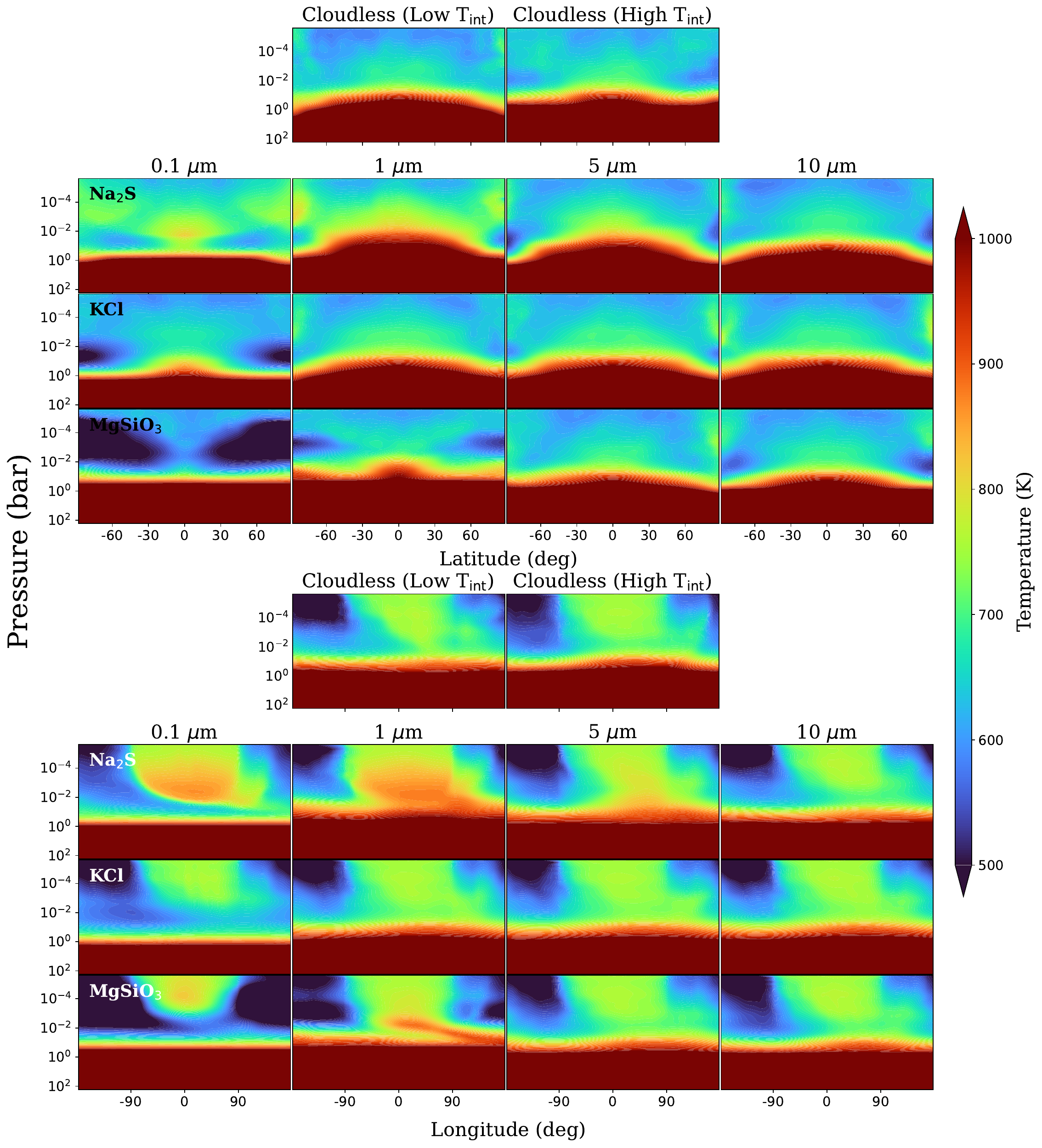}
    \caption{Top: Latitudinal distribution of temperature (shows equatorial and polar regions), averaged over longitudes for all models.  Within each subplot, the x-axis shows the latitude and the y-axis the pressure.  The first row shows cloudless models, low (100 K) and high (381 K) T$_{\rm int}$. Rows 2, 3, and 4 correspond to Na$_2$S, KCl, and MgSiO$_3$ (with high T$_{\rm int}$) clouds, respectively, with each column representing a different particle size indicated at the top. Bottom: Same as top, but for longitudinal distribution of clouds (shows dayside and nightside of the planet), averaged over latitudes for all models.}
    \label{fig:ptmean}
\end{figure*}

The cloud distribution on the planet is determined by both the thermal structure and the dynamical transport. The subplots of Fig. \ref{fig:pclds} show the cloud distribution averaged over latitude and longitude. In Fig. \ref{fig:ptracer}, the globally averaged mass mixing ratios of condensate and vapour are compared to the saturation mixing ratio.
Comparing the cloud distribution (Fig. \ref{fig:pclds}) with the temperature structure (Fig. \ref{fig:ptmean}), it can be seen that the clouds are present in the cooler regions of the planet. In Fig. \ref{fig:pclds}, every row shows a clear trend of cloud distribution with the particle size. Smaller particle-size clouds are lofted high in the atmosphere and are evenly distributed over the planet. As the particle size increases, the clouds settle down to the higher pressures and start forming clusters in the cooler regions of the planet. The clouds are depleted around the super-rotating jet and the dayside. The temperature shapes the cloud boundary in the deep atmosphere, but the winds shape the upper atmosphere. 

As seen in Fig. \ref{fig:pclds}, the Na$_2$S clouds with particle size of 0.1 $\mu$m are evenly distributed over the latitude and longitude between the pressures 0.01 bar to 1 bar. At lower pressures, the clouds are depleted near the equator due to a super-rotating jet, along with clouds being evaporated on the dayside. This is observed by the asymmetrical distribution due to the eastern winds. The distribution of 0.1 $\mu$m Na$_2$S clouds is similar to the distribution of the temperature. The clouds are present in the cooler regions of the planet and are lofted high in the atmosphere. The 1 $\mu$m Na$_2$S clouds are dense at the poles at higher pressure (0.1 - 1 bar) while scarce near the equator, dayside, and lower pressures. The eastern limb is less cloudy than the western limb due to the eastward transport by the equatorial jet. 
With the larger particle size, the clouds settle down at high pressures and concentrate near the poles. Hence, they are not evenly distributed over the planet, but according to the temperature structure. The cloud distribution of the 5 $\mu$m cloud is very uneven. The clouds are clustered near the poles with scarce distribution near the equator, and evenly distributed over the day and night. These clouds have settled down to the higher pressures and are only present in the cooler regions of the planet. The GCM with 10 $\mu$m Na$_2$S clouds is depleted throughout the planet. The clouds are mainly distributed near the poles, covering all longitudes. Due to the large particle size, even in cooler regions at lower pressures, the clouds have settled down to higher pressures.

We now turn to the KCl clouds. KCl has both a lower abundance and a cooler condensation temperature. However, similar to the Na$_2$S case, they can still evaporate at the low pressures of the dayside. 
The 0.1 $\mu$m KCl clouds are mainly present in pressures between 1-0.01 bar, and these clouds are depleted in the equatorial region.
As shown in Fig. \ref{fig:pclds}, 0.1 $\mu$m KCl clouds are evenly spread over the longitudes. 
Whereas the 1 $\mu$m KCl clouds are present on the nightside and evaporate from the dayside. A shift can be seen in the longitudinal distribution due to the strong super-rotating jet and the prograde winds.
But for 5 and 10 $\mu$m KCl clouds are only present near the polar region and hence over all the longitudes. The increase in size from 5 to 10 $\mu$m only increases the cluster size at the polar regions. The KCl clouds are formed in the cooler regions of the planet. The smaller particles are lofted and hence evenly distributed over the planet. The larger particles settle down, deviating from the temperature structure of the planet, and start forming clusters in the atmosphere. 

Finally, the last two rows of Fig. \ref{fig:pclds} show the distribution of MgSiO$_3$ clouds. These clouds have a much higher condensation temperature and, as such, condense in the deep layers of the planet, around 10 bars.
At pressures lower than 10 bar, the cloud distribution is mainly influenced by dynamical mixing rather than the local thermal structure. As a result, small particles (0.1 and 1 $\mu$m) are efficiently lofted to higher altitudes. The clouds are relatively homogeneous across the dayside and nightside; however, a depletion is observed near the equator, caused by the strong equatorial super-rotating jet. 1 $\mu$m MgSiO$_3$ settle down in the atmosphere and are only present in the 1 - 0.001 bar pressure region.
In contrast, larger particles (5 or 10 $\mu$m) remain confined below the photospheric layers around 10 bar. 5 $\mu$m clouds are scarcely distributed at lower pressures between 1 - 0.001 bar.

\subsection{Effect of clouds on temperature}
\label{section: cldsontemp}
Overall, the photospheric temperature structure of the planet is very homogeneous with a very efficient heat redistribution for the cloudless case ($f$=1.04). Adding different clouds to the planet causes changes in the temperature structure depending on the species, particle size, and abundance of the clouds. 

The pressure-temperature profile of cloudless GCMs is shown in Fig. \ref{fig:ptp_comb}. Different clouds affect the temperature structure depending on the species, particle size, abundance, and location of the clouds. The condensation curves of the Na$_2$S, KCl, and MgSiO$_3$ clouds are shown in Fig. \ref{fig:ptp_comb} along with the pressure-temperature profile of the cloudless GCM. The pressure-temperature profile of the cloudless GCM crosses the condensation curves of Na$_2$S and KCl clouds, whereas the pressure-temperature profile does not cross the condensation curve of MgSiO$_3$ clouds. 
As a consequence, for our low T$_{\rm int}$ model, only Na$_2$S and KCl clouds can be present in the modelled atmosphere. In order to study the possible distribution and impact of MgSiO$_3$ on the spectra, we further simulate models with hotter deep thermal profiles, corresponding to a higher T$_{\rm int}$ of 381 K as proposed by \cite{wiser2025}. 

In the 0.1 $\mu$m Na$_2$S cloud case, the planet’s temperature structure is affected significantly.
The 0.1 micron Na$_2$S clouds have a strong radiative feedback on the atmosphere. This feedback depends on planetary location: on the dayside and eastern terminator, they create a strong thermal inversion at pressure less than 0.02 bar and cool down the layers below it. On the nightside and the western terminator, they lead to a cooler atmosphere. The thermal inversion also occurs on the western terminator, but at a lower pressure of 0.001 bar. Due to the high Mass Mixing Ratio, the 0.1 $\mu$m Na$_2$S particles are very efficient in absorbing the stellar radiation, causing the temperature inversion above the clouds. They shield the atmosphere below them from the stellar radiation and hence cool the atmosphere below them. The inversion is stronger on the dayside as the stellar radiation is more intense on the dayside. The nightside is cooler because the dayside is cooler at pressures greater than 0.02 bars. So less heat is transported to the nightside.  
1 $\mu$m Na$_2$S clouds absorb across the spectrum, leading to strong heating in the location of the clouds. Hence, a heating effect is seen in the upper atmosphere (above 1 bar) where the clouds are present. There is no cooling effect as these clouds do not absorb the stellar radiation as efficiently as the 0.1 $\mu$m Na$_2$S clouds. The 5 $\mu$m Na$_2$S clouds do not absorb radiation efficiently and hence do not affect the temperature structure of the planet significantly; they slightly increase the temperature in the upper atmosphere. The 10 $\mu$m Na$_2$S clouds mostly follow the temperature structure of the cloudless GCM due to lower abundance and not significant absorption of radiation. 

KCl clouds are less abundant in the atmosphere; only 0.1 $\mu$m KCl clouds affect the temperature significantly due to the optical properties (see Fig. \ref{fig:ssa}). Clouds with larger particle sizes do not affect the temperature structure of the planet significantly. 
From Fig. \ref{fig:ptp_comb}, we see that the condensation curve crosses the cloudless dayside average pressure-temperature profile between 0.002 and 0.05 bar. 
In the case of 0.1 $\mu$m KCl clouds, a net cooling of the atmosphere in the GCM occurs despite their relatively low abundance. This cooling occurs without the formation of a thermal inversion, indicating that shortwave absorption by KCl particles is weak. Instead, sub-micron KCl clouds primarily scatter and reflect incident stellar radiation, increasing the planetary albedo and reducing the net stellar energy absorbed by the atmosphere. As a consequence, the atmosphere cools throughout the irradiated layers without heating above the cloud deck.
In contrast, 0.1 $\mu$m Na$_2$S clouds, which are more abundant, exhibit significant short-wave absorption that results in strong thermal inversions through localised heating at low pressures. This difference in atmospheric response primarily reflects the distinct optical properties of the two condensates, with the higher abundance of Na$_2$S further amplifying its absorption-driven radiative impact.
When larger KCl clouds are included, they raise the temperature in this region. As a result, the dayside average pressure-temperature profile no longer crosses the condensation curve, meaning that KCl is unlikely to condense there. However, the condensation curve still intersects the pressure-temperature profiles from other regions of the planet, where cloud formation may still occur. These clouds are present in the cooler regions of the planet and absorb radiation efficiently, hence slightly cooling the atmosphere below them while heating the upper atmosphere. The change in the temperature structure is not significant.

With the high T$_{\rm int}$, the pressure-temperature profile crosses the condensation curve of MgSiO$_3$ clouds. 
The 0.1 $\mu$m MgSiO$_3$ clouds are present all over the planet. With a small particle size, these clouds are lofted high in the atmosphere. Their radiative behaviour is similar to that of 0.1 $\mu$m Na$_2$S clouds: both absorb stellar radiation efficiently at low pressures, leading to a temperature inversion on the dayside around 0.01 bar. On the eastern, western, and nightsides, the inversion occurs at higher altitudes due to reduced stellar heating.
On the other hand, the deeper atmosphere is warmer due to the presence of clouds with higher absorption opacity, in contrast to the 0.1 $\mu$m Na$_2$S case. This difference arises from the distinct optical properties of the clouds, as shown in Fig. \ref{fig:ssa}.
The 1 $\mu$m MgSiO$_3$ clouds increase the atmospheric temperature below 1 bar, causing the T-P profile to overlap with the condensation curve in this region. This creates an interesting scenario in which clouds continuously form and evaporate due to radiative feedback. Initially, when the T-P profile is cooler than the condensation curve, cloud condensation occurs. The resulting clouds then warm the local atmosphere through radiative heating, raising the temperature above the condensation curve and leading to cloud evaporation. This cycle of condensation and evaporation can repeat continuously and introduces numerical instabilities in the GCM. In addition, the 1 $\mu$m MgSiO$_3$ clouds produce a temperature inversion around 0.01 bar.
5 and 10 $\mu$m MgSiO$_3$ clouds are present deep in the atmosphere and do not affect the temperature structure of the planet.

\subsection{Effect of clouds on dynamics}

Clouds alter the thermal distribution, which in turn affects atmospheric dynamics and transport by influencing the heating and cooling of the atmosphere. Different cloud species influence the atmospheric wind structure in distinct ways, as shown in Fig. \ref{fig:puwmean}. The zonal, meridional, and vertical wind components, each averaged over latitude and longitude, are presented in Appendix Figs. \ref{fig:pumean}, \ref{fig:pvmean}, and \ref{fig:pwmean}, respectively. In Fig. \ref{fig:puwmean}, the 0.1 $\mu$m Na$_2$S cloud case stands out relative to the cloudless model, it develops a strong equatorial jet accompanied by retrograde winds at high latitudes. Similar features appear in the 0.1 $\mu$m and 1 $\mu$m MgSiO$_3$ cloud simulations, both of which were run with a high $T_{\rm int}$. Notably, the circulation patterns in these three cloudy cases strongly resemble those of the cloudless high-$T_{\rm int}$ model. This suggests that the 0.1 $\mu$m Na$_2$S clouds significantly affect the dynamics, whereas the 0.1 and 1 $\mu$m MgSiO$_3$ clouds have comparatively little dynamical impact.

The strong influence of the 0.1 $\mu$m Na$_2$S clouds likely arises from the temperature inversion they induce. The increase in temperature with altitude above $\sim$0.1 bar can create enhanced dayside upwelling and also produce the observed high-latitude flow reversal, similar to that found in the high-$T_{\rm int}$ cloudless case.

More broadly, our simulations illustrate the strength of upwelling. They also show whether retrograde high-latitude winds form or whether prograde flow remains dominant across the planet. The 0.1 $\mu$m Na$_2$S clouds induce substantial dynamical changes, leading to retrograde polar flow in contrast to the globally prograde circulation of the corresponding low-$T_{\rm int}$ cloudless model. In comparison, the 5 and 10 $\mu$m MgSiO$_3$ clouds also modify the dynamics but ultimately produce wind structures that closely resemble those of the cloudless low-$T_{\rm int}$ case, despite being modelled with high $T_{\rm int}$. Additionally, the 1 and 5 $\mu$m Na$_2$S clouds significantly weaken, or nearly eliminate, the equatorial jet. For all remaining cloud cases, the dynamics largely follow that of their respective cloudless models, with variations in jet strength depending on cloud species.
Appendix Fig. \ref{fig:jetspeed} summarises how the jet speed (maximum zonal-mean zonal wind) varies with particle size for each of the three cloud species.

\begin{figure*}
    \centering
    \includegraphics[width=\textwidth]{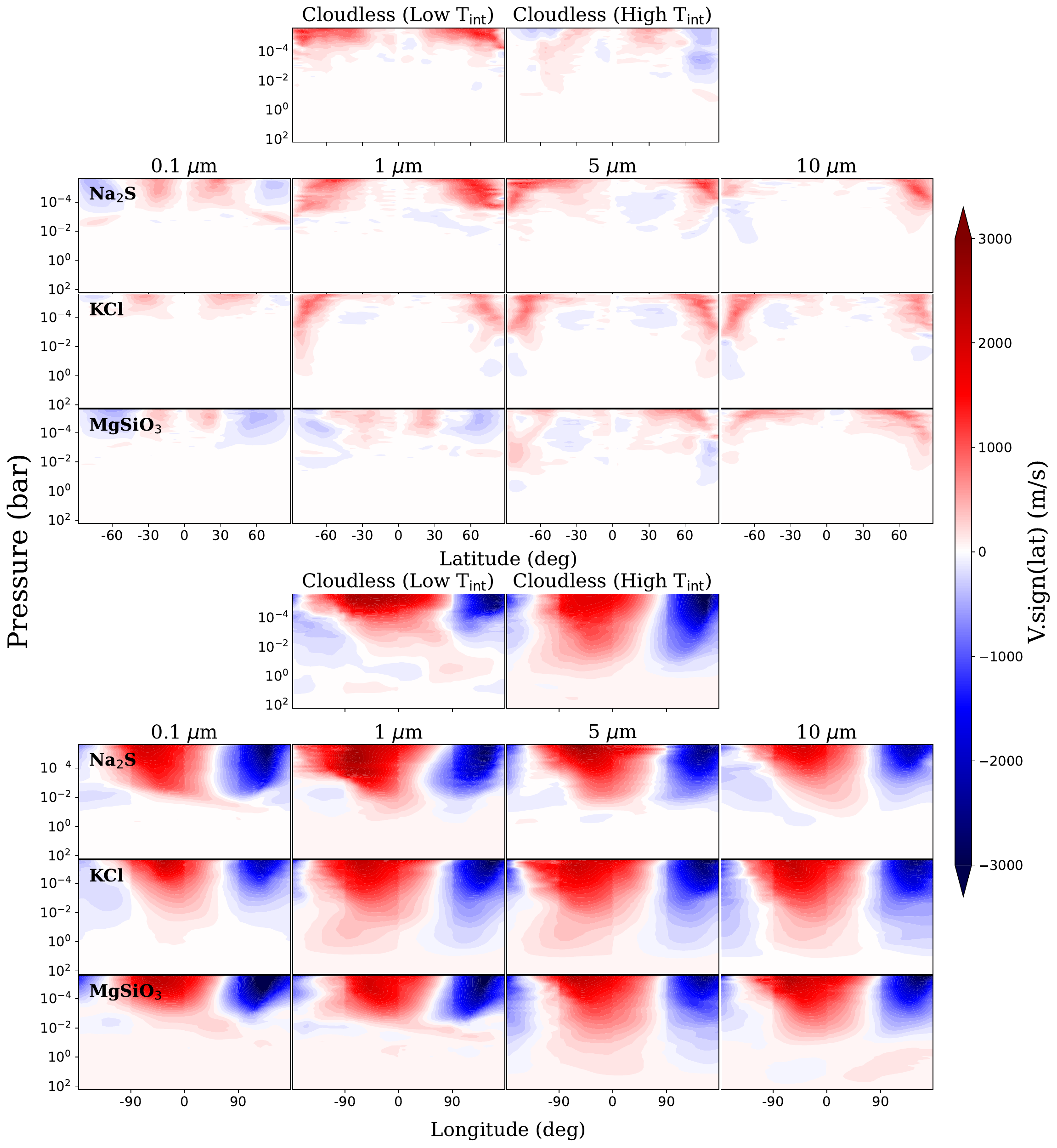}
    \caption{Same as Fig. \ref{fig:pclds} but for meridional wind multiplied by the sign of latitude to indicate wind towards (negative) or away (positive) from the equator. }    
    \label{fig:pvsign}
\end{figure*}

\begin{figure}
    \centering
    \includegraphics[width=\columnwidth]{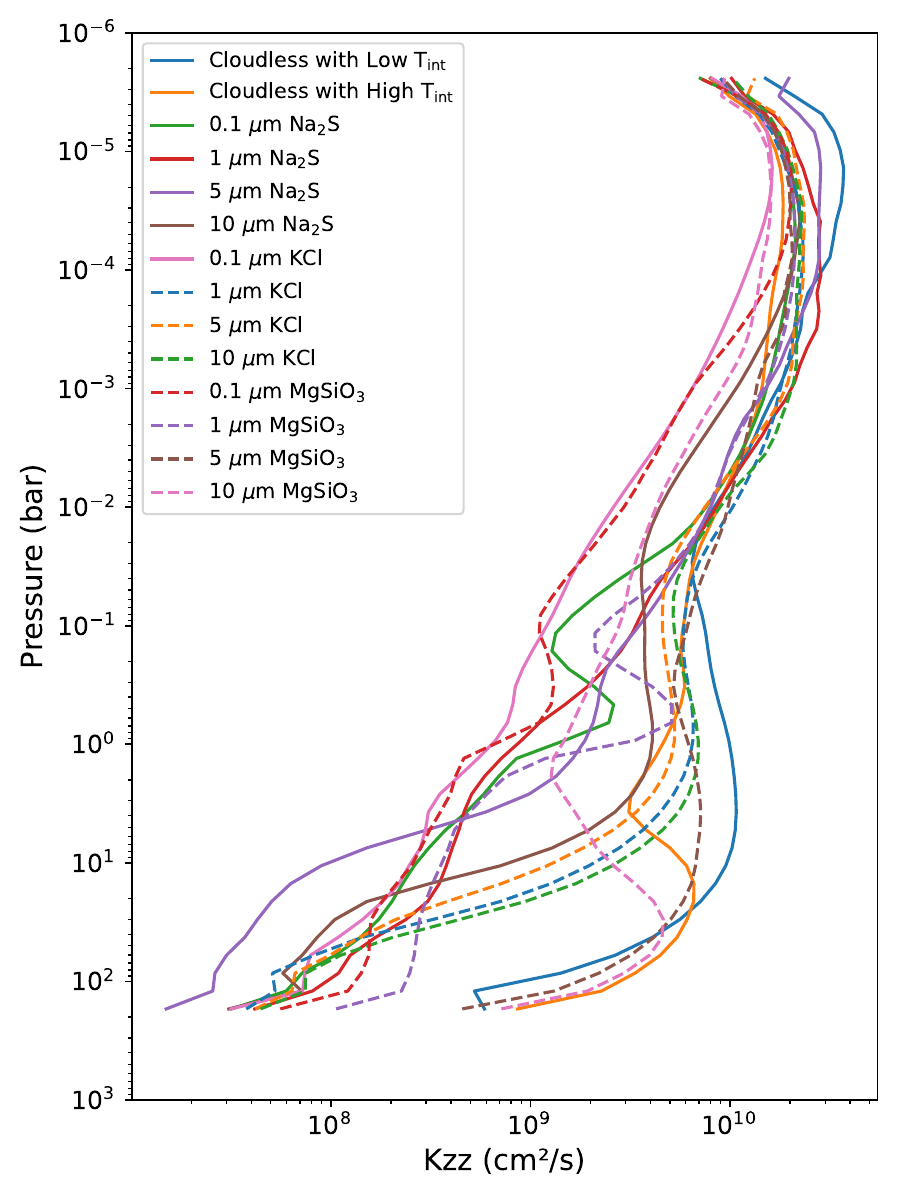}
    \caption{K$_{zz}$ profiles for different cloud cases. K$_{zz}$ was calculated as the root mean square of the vertical velocity times the vertical scale height.}
    \label{fig:pkzz}
\end{figure}

Fig. \ref{fig:pvsign} presents the meridional wind component multiplied by the sign of latitude, effectively illustrating the direction of atmospheric flow with respect to the equator. This projection highlights regions of poleward and equatorward transport. From Fig. \ref{fig:pvsign}, we can see that the winds and their direction have a similar distribution over longitude, but the depth of these winds that move towards the equator or poles and their strength differ with different cloud cases. Fig. \ref{fig:pkzz} shows the K$_{zz}$ for all GCM cases. The K$_{zz}$ profiles for different cloud cases are calculated by $K_{zz}(p)\sim H\centerdot w_{\rm rms}(p)$, where H is the scale height, and w is the vertical component of winds as a function of pressure \citep{lewis2010, moses2011}. Although this approach usually overestimates the mixing strength by one or two orders of magnitude \citep{parmentier2013a}, it is useful in our case to compare the relative mixing efficiency of the different models. We see that all models have approximately the same mixing strength at pressures lower than 0.1 bar. However, the presence of clouds tends to reduce the mixing strength in the deep layers of the atmosphere, with some cloud species, such as Na$_2$S and KCl, reducing the mixing by 2 orders of magnitude.

\subsection{Comparison with the observations}
\label{subsection: spectrum}

\begin{figure*}
    \centering
    \includegraphics[width=\textwidth]{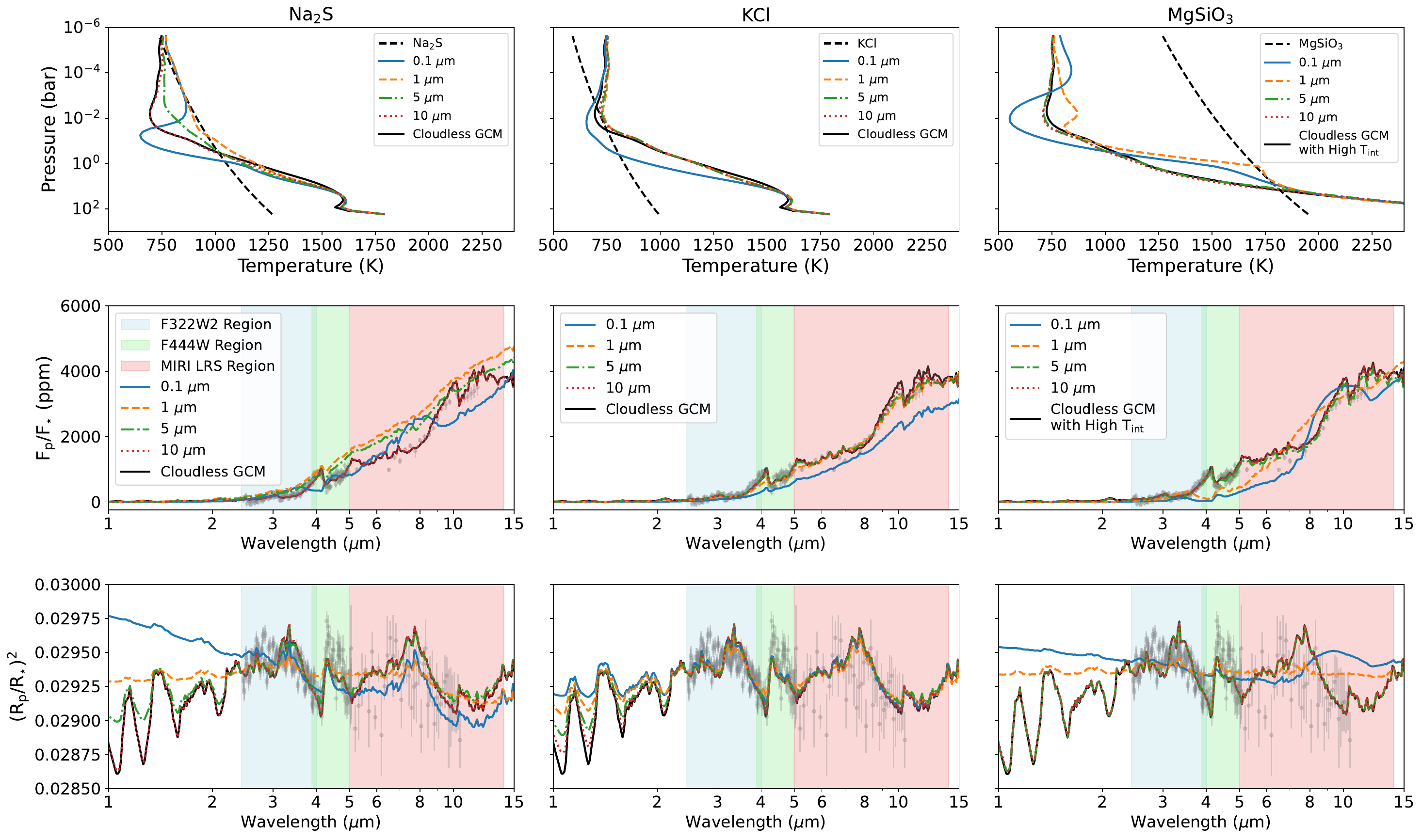}
    \caption{Top row: Dayside-averaged pressure-temperature profiles for different particle sizes along with the cloudless case (solid black line). The condensation curve for the corresponding cloud species is plotted with a black line. Middle row: Emission spectra from GCM models along with the cloudless case (solid black line) and JWST observation (grey). Bottom row: Transmission spectra from GCM models along with the cloudless case (solid black line) and JWST observation (grey).}
    \label{fig:pspec}
\end{figure*}

The transmission and emission spectra were obtained for cloudy GCMs with different particle sizes of Na$_2$S, KCl, and MgSiO$_3$ clouds. The transmission and emission spectra were compared with the observations from JWST. The transmission and emission spectra are shown in Fig. \ref{fig:pspec}. The nightside temperature-pressure profiles and emission spectra are shown in Appendix Fig. \ref{fig:pnight}.

Condensation clouds have been shown to exist on numerous planets, and we expect them to form in WASP-80b. Particularly, as shown in Fig. \ref{fig:ptp_comb}, the thermal profile of the planet crosses the condensation curves of Na$_2$S, KCl near the photosphere and, for the high T$_{\rm int}$ scenarios, the condensation curve of MgSiO$_3$ at depth. As a consequence, even though both transmission and emission spectra are compatible with the cloudless model, we do expect clouds to form in this atmosphere. The WASP-80b dataset, therefore, provides us with an opportunity to determine which kind of clouds can form and which ones are unlikely to form. For this, we run models with three different types of clouds, and, for each of them, 4 different mean particle sizes. Each of these clouds can have an indirect effect on the spectra by changing the thermal structure of the planet and a direct effect by spectral signatures in the emission and transmission spectra.

For each GCM scenario, we generated the emission and transmission spectra and compared them against the observations, which enabled us to place stronger constraints on the cloud properties of the planet. The $\chi^2_{\rm red}$ was calculated for each model comparing the transmission and emission spectra with the observations. The chi-square was used to rule out the models that are not able to fit the observations.

We note that 10 $\mu$m Na$_2$S clouds provide a good match for emission and transmission, as it is similar to the cloudless case (Section \ref{section: cloudless}).
As seen in Fig. \ref{fig:pclds} and discussed in Section \ref{subsection: spatial cloud distribution}, 10 $\mu$m Na$_2$S clouds tend to concentrate at pressures of 0.1 bar and near the poles, making them both invisible to the spectra and having a weak impact on the radiative transfer. Whereas, the models with 0.1, 1, and 5 $\mu$m Na$_2$S clouds are all ruled out by the data; this is mainly due to their effect on the dayside thermal profile. Indeed, the presence of 0.1 $\mu$m clouds leads to a strong thermal inversion on the dayside atmosphere, leading to emission features rather than absorption, which is incompatible with the observations. For the 1 and 5 $\mu$m cases, the dayside thermal profile is not inverted; however, the greenhouse effect of the clouds on the nightside atmosphere leads to a hotter and more isothermal dayside. Thus, despite the lack of strong cloud coverage on the dayside, the emission spectrum becomes too large and the spectral features too small to be compatible with the observations. 

For the transmission spectra, the 0.1, 5, and 10 $\mu$m Na$_2$S clouds can fit the data. The 0.1 $\mu$m Na$_2$S clouds can fit the terminator data. These clouds lift the baseline of the transmission spectrum at lower wavelength regions (<2.5 $\mu$m) where the data is unavailable (NIRISS/SOSS: This observation will be part of the JWST-GO-5924 programme; PI: David Sing; NIRCam F210M: The data was not utilised due to excess noise). 5 $\mu$m Na$_2$S clouds also lift the baseline of the transmission spectrum at lower wavelength regions, but not as much as the 0.1 $\mu$m Na$_2$S clouds. This is due to the distribution of clouds, where the smaller particles are present in the upper atmosphere, leading to a higher baseline. As the size increases, the changes in the baseline decrease. The 10 $\mu$m Na$_2$S clouds are present in the deep atmosphere and do not affect the transmission spectrum. On the other hand, the 1 $\mu$m Na$_2$S clouds suppress all the features in the transmission spectrum and fail to reproduce the CO$_2$ feature at 4.3 $\mu$m. This is because these clouds are present in the upper atmosphere, affecting the baseline of the transmission spectrum. Although the spatial distribution of 0.1 and 1 $\mu$m Na$_2$S clouds is similar at lower pressures, the 0.1 $\mu$m particles provide a better fit to the observations. This is because their absorption cross-section is higher at shorter wavelengths, allowing spectral features at longer wavelengths to be unaffected. In contrast, the 1 $\mu$m particles exhibit significant absorption across the entire wavelength range, resulting in a broad suppression of spectral features and a poorer match with the data.

For 1, 5, and 10 $\mu$m KCl clouds, the models can fit the dayside emission spectrum. These clouds do not affect the emission spectrum significantly since they do not affect the temperature structure of the planet significantly due to their low optical depth \citep{fortney2005,lee2025} and low abundance. Whereas 0.1 $\mu$m KCl clouds are not able to fit the emission spectrum, as the dayside flux is underestimated due to the lower temperature caused by the clouds.
KCl clouds affect the transmission spectrum at wavelengths below 2.5 $\mu$m by raising the baseline of the spectrum, with the magnitude of this effect depending on particle size. As the particle size increases, the clouds settle down to the higher pressures, and the transmission spectrum is less affected by the clouds. The baseline of the transmission spectrum is elevated at lower wavelengths, primarily by 0.1 $\mu$m KCl clouds, and it is unaffected by the 10 $\mu$m KCl clouds.

The MgSiO$_3$ clouds were included in the GCMs with high T$_{\rm int}$. As seen in Fig. \ref{fig:pspec}, the 5 and 10 $\mu$m MgSiO$_3$ clouds have the best fit to observations according to Table \ref{tab:chi-sq}.
The 5 and 10 $\mu$m MgSiO$_3$ clouds do not affect the dayside emission spectrum significantly due to the clouds settling in deeper layers of the atmosphere. On the other hand, the dayside spectra of 0.1 and 1 $\mu$m MgSiO$_3$ clouds show emission features instead of absorption features due to the thermal inversion. This is incompatible with the observations.
Even for the transmission spectrum, the models with 5 and 10 $\mu$m MgSiO$_3$ clouds can fit the terminator observations. The 0.1 and 1 $\mu$m MgSiO$_3$ clouds lift the baseline of the transmission spectrum, leading to a near-flat transmission spectrum. A spectral feature is seen at 10 $\mu$m in the 0.1 $\mu$m MgSiO$_3$ cloud case, which is due to the absorption of the clouds. 

According to the Emission models from Table \ref{tab:chi-sq}, we see that smaller particle size Na$_2$S, KCl, and MgSiO$_3$ clouds have a much higher $\chi^2_{\rm red}$value than other models. Remaining models have comparable $\chi^2_{\rm red}$ values, making it difficult to choose one correct model, but we can say with certainty that the clouds rejected by these models (smaller particle size Na$_2$S, KCl, and MgSiO$_3$ clouds) should not be present on the planet. On the other hand, the majority of the transmission models have similar $\chi^2_{\rm red}$. We can reject the 1 $\mu$m Na$_2$S, 0.1 and 1 $\mu$m MgSiO$_3$ model based on the transmission spectrum; while the transmission spectrum is usually more sensitive to clouds \citep{fortney2005, inglis2024,welbanks2024}, we find the opposite for our models of WASP-80b. 
Overall, these results demonstrate that a combined analysis of transmission and emission spectra is required to robustly characterise clouds on warm Jupiters, as interpretations based on a single spectrum may be misleading.

\section{Discussions and conclusion}
\label{section: conclusion}

We modelled the three-dimensional atmospheric structure of the warm Jupiter WASP-80b, both with and without clouds, and compared the results to JWST emission and transmission spectra. Our model suite included cloud-free cases with low and high internal heat flux (T$_{\rm int}$), as well as cloudy cases with various mean particle radii (0.1, 1, 5, and 10 $\mu$m) and different condensate species (Na$_2$S, KCl, and MgSiO$_3$), treated as radiatively active tracers. This grid of models allowed us to investigate how the thermal structure and atmospheric dynamics respond to different cloud properties and internal heat flux scenarios.

The spatial distribution of clouds depends on the local pressure, local temperature, dynamics, and the condensation properties of individual species. We observe the similarity between the cloud distribution (Fig. \ref{fig:pclds}) and the temperature distribution (Fig. \ref{fig:ptmean}). The dynamics of the planet are an additional factor that comes into play. In addition, dynamics play a key role: the balance between vertical winds and particle settling velocities sets the vertical extent of the cloud deck, while horizontal winds shape the global distribution, producing asymmetries between the planetary limbs. Such limb asymmetries have already been detected in JWST observations of several exoplanets \citep{espinoza2024, murphy2024, mukherjee2025}. Moreover, radiative feedback from clouds amplifies temperature contrasts between the limbs, as demonstrated by \citet{murphy2025} using the same GCM framework.

Every cloud species and particle size has different optical properties and optical depth, which in turn determines the strength of their radiative feedback. Na$_2$S and MgSiO$_3$ clouds generally have a strong feedback, while KCl clouds, being optically thin and less abundant, have a weak radiative feedback. The radiative feedback also depends on the abundance of the clouds and their spatial distribution. 
Cloud coverage on the dayside in particular can significantly alter the atmospheric dynamics. The 0.1 $\mu$m Na$_2$S, 0.1 $\mu$m MgSiO$_3$, 5 $\mu$m MgSiO$_3$ clouds had a major impact on the zonal-mean zonal wind structure, whereas the other cloud cases resulted in wind patterns similar to their respective cloud-free simulations with low or high tint. The changes in the wind structure are due to the radiative feedback of the different cloud cases. 
The effect of radiative feedback on dynamics can be further seen in the vertical and meridional wind components (and consequently in the K$_{zz}$ profiles; see Figs. \ref{fig:pvmean} and \ref{fig:pwmean}). This shows the importance of radiative feedback of clouds on the dynamics of warm giant planets, especially when the clouds are present on the dayside. In atmospheres with highly efficient heat redistribution ($f$ = 1.04, calculated from the cloudless case), even small perturbations to the thermal structure can drive noticeable dynamical changes. In addition to clouds, the effective internal temperature also strongly modulates atmospheric dynamics, consistent with \cite{komacek2022}.

As shown in Section \ref{section: cloudless}, our cloud-free model exhibits very good agreement with both the emission and transmission spectra of WASP-80b, without requiring any additional tuning beyond adopting the chemical abundances retrieved from the 1D spectral analysis and setting the reference pressure in post-processing. This result demonstrates that GCMs are capable of capturing most of the physics governing the three-dimensional thermal structure of this warm giant planet, thereby providing a robust benchmark for GCM studies. The good agreement with observations is surprising, as the temperature profile of WASP-80b crosses multiple condensation curves.

Models including larger KCl clouds also reproduce the observations with high accuracy. There are several reasons: (i) the KCl condensation curve only marginally intersects the dayside temperature–pressure profile (see Fig. \ref{fig:pspec}), with the majority of cloud formation occurring on the nightside (see Fig. \ref{fig:pnight}); (ii) the elemental abundances of K and Cl are relatively low; and (iii) KCl clouds are optically thin and less abundant. As a result, their impact on the dayside emission spectrum is negligible and only minor in the transmission spectrum. Together, these effects cause the spectra of larger particle-sized KCl-cloud models to closely resemble those of the cloud-free case, even when clouds are present.

Models with large cloud particle sizes can reproduce the spectra. In these cases, efficient gravitational settling removes clouds from the observable atmosphere, minimising their spectral impact. \cite{powell2024} have shown that cloud particle distributions can extend up to 100 $\mu$m; however, particles larger than $\sim$10 $\mu$m would not significantly affect the spectra. 

From this study, we were able to rule out certain cloud species that are not consistent with the JWST observations. 0.1 $\mu$m Na$_2$S clouds are optically thick and cause a strong radiative feedback, leading to a temperature inversion. Larger Na$_2$S particles (1 and 5 $\mu$m) instead raise the photospheric temperature, resulting in an overestimation of the emitted flux and the emission spectrum. Indeed, it remains debated whether Na$_2$S clouds can form in exoplanet atmospheres, as their high surface energies create a substantial barrier to nucleation \citep{gao2020}. 

Finally, we note that certain clouds (e.g. 0.1 $\mu$m Na$_2$S) can reproduce the dayside observations but cannot match the terminator spectra. This demonstrates the importance of multi-view observations in constraining exoplanet atmospheres, as dayside and limb spectra provide different, complementary information.

From our analysis, we draw the following main conclusions:
\begin{enumerate}
    \item WASP-80b is a warm Jupiter with a very efficient heat redistribution. The thermal distribution of the planet is very homogeneous, with the possibility that the planet may be cloudless. 
    \item The cloudless GCM was able to reproduce the observations of the planet with a good fit to the dayside and terminator observations. The relatively low abundance of CH$_4$ on the terminator can be due to photochemistry. 
    \item Large particle clouds of any species can be present as they settle down in the atmosphere and do not affect the spectra significantly. 
    \item KCl clouds ($> 0.1~\mu$m) can be present on the planet as they do not cause a significant change in transmission or emission spectrum for this planet. 
    \item Cloud particles smaller than the identified thresholds, Na$_2$S ($<10~\mu$m), KCl ($<1~\mu$m), and MgSiO$_3$ ($<5~\mu$m), are inconsistent with the JWST observations due to their strong radiative feedback.

\end{enumerate}
Overall, we show that multiple cloudy scenarios are compatible with the apparently cloudless spectrum of WASP-80b. We demonstrate that observations at short wavelengths should break the degeneracy between these scenarios and determine whether WASP-80b is truly cloudless or whether clouds exist but are not currently affecting the observations. 

The possibility of the presence of PH$_3$ and CS$_2$ in the atmosphere of WASP-80b is out of scope for this study, but the chemistry for these molecules should be further explored. A recent study \citep{veillet2025} showed that there is an increase in the production of CS$_2$ on a planet with a temperature of around 800 K with an updated chemical network. A detailed study on the presence of PH$_3$ and CS$_2$ on WASP-80b should be explored in detail. 

We can enhance this framework and impose tighter constraints on cloud properties by incorporating more realistic assumptions from cloud microphysics, particularly regarding cloud species and particle sizes. A dedicated microphysical study, similar to that of \cite{powell2024}, focused on atmospheres with equilibrium temperatures around 800 K, could inform better assumptions for particle size distributions, especially since different condensates have been shown to produce distinct particle sizes. We can also increase the complexity by including multiple cloud species simultaneously. This would lead to complex effects on the thermal structure. 

WASP-69b (\(T_{\rm eq} \sim 963\,\mathrm{K}\)), WASP-80b (\(T_{\mathrm{eq}} \sim 820\,\mathrm{K}\)), and WASP-107b (\(T_{\mathrm{eq}} \sim 735\,\mathrm{K}\)) are planets of similar size, orbiting stars of the similar spectral type, and lie within the same equilibrium temperature range. Yet, their atmospheres exhibit different characteristics. WASP-69b and WASP-80b show clear differences in heat redistribution efficiency, and clouds are favoured in models to fit the JWST spectrum of WASP-69b, either through high albedo or by creating a highly inhomogeneous dayside \citep{schlawin2024}. In contrast, WASP-80b shows no strong observational indications of clouds in its atmosphere. On the other hand, the JWST transmission spectrum of WASP-107b strongly supports the presence of a cloud deck \citep{welbanks2024, sing2024} as well as strong limb asymmetry \citep{murphy2025}. A detailed comparative study of these planets, within this shared temperature regime, using GCMs coupled with clouds, is necessary to understand the origins of their atmospheric diversity better (Mehta et al., in prep.).

\begin{acknowledgements}

This work was supported by the French government through the France 2030 investment plan managed by the National Research Agency (ANR), as part of the Initiative of Excellence Université Côte d’Azur under reference number ANR-15-IDEX-01. The authors are grateful to the Université Côte d’Azur’s Center for High-Performance Computing (OPAL infrastructure) for providing resources and support.\\

The JWST data presented in this article were obtained from the Mikulski Archive for Space Telescopes (MAST) at the Space Telescope Science Institute. The specific observations analysed can be accessed via DOI: \href{https://doi.org/10.17909/ms85-4a43}{10.17909/ms85-4a43}.
\end{acknowledgements}

\bibliographystyle{aa}
\bibliography{aa57411-25}

\newpage

\begin{appendix}
\label{appendix}
\section{Supplementary information of atmospheric structure and cloud diagnostics}

\begin{figure*}[b]
    \centering
    \includegraphics[width=\textwidth]{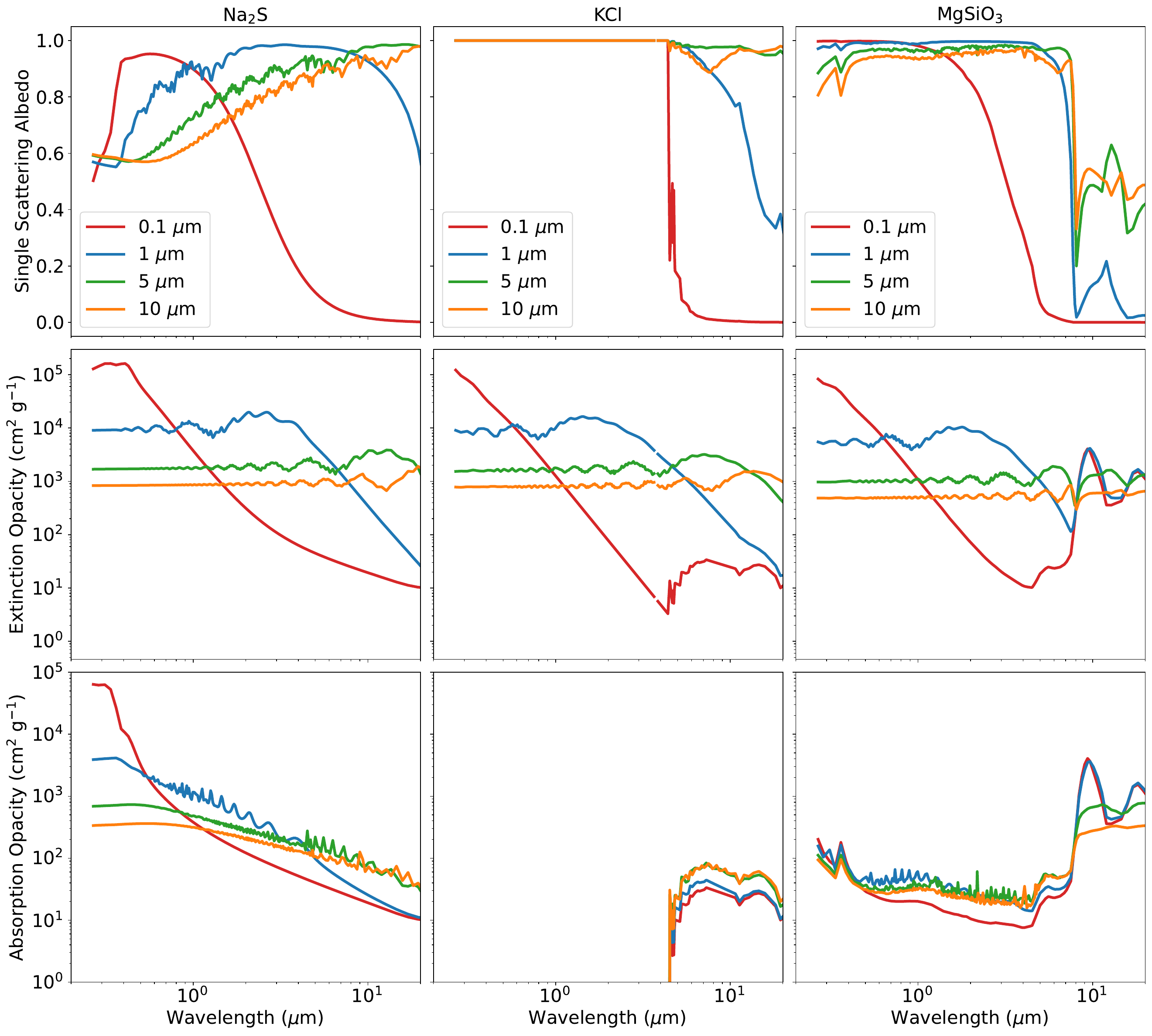}
    \caption{Wavelength-dependent single scattering albedo (top row), extinction opacity (middle row), and absorption opacity (bottom row) for Na$_2$S (left), KCl (middle), and MgSiO$_3$ (right) clouds. They are calculated for particle radii of 0.1, 1.0, 5.0, and 10.0 $\mu$m using \texttt{PyMieScatt} \citep{sumlin2018}.}
    \label{fig:ssa}
\end{figure*}

\begin{figure*}
    \centering
    \includegraphics[width=\textwidth]{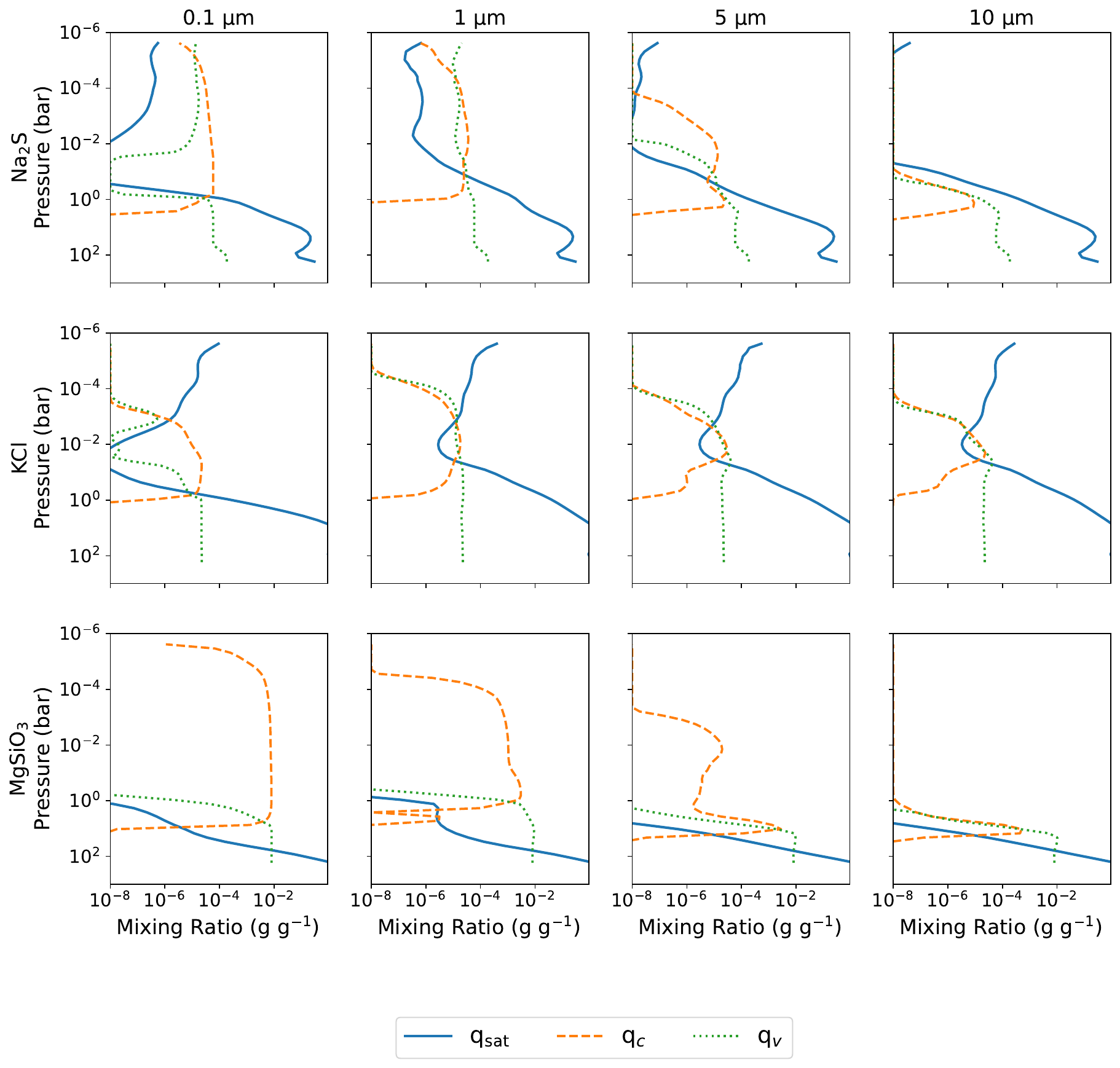}
    \caption{Globally averaged mass mixing of condensate (q$_c$) and vapour (q$_v$) compared with saturation mixing ratio (q$_{\rm sat}$). The columns show the radii of the cloud particle (0.1, 1, 5, 10 $\mu$m) and the rows show the cloud species (Na$_2$S, KCl, MgSiO$_3$)}
    \label{fig:ptracer}
\end{figure*}

\begin{figure*}
    \centering
    \includegraphics[width=0.93\textwidth]{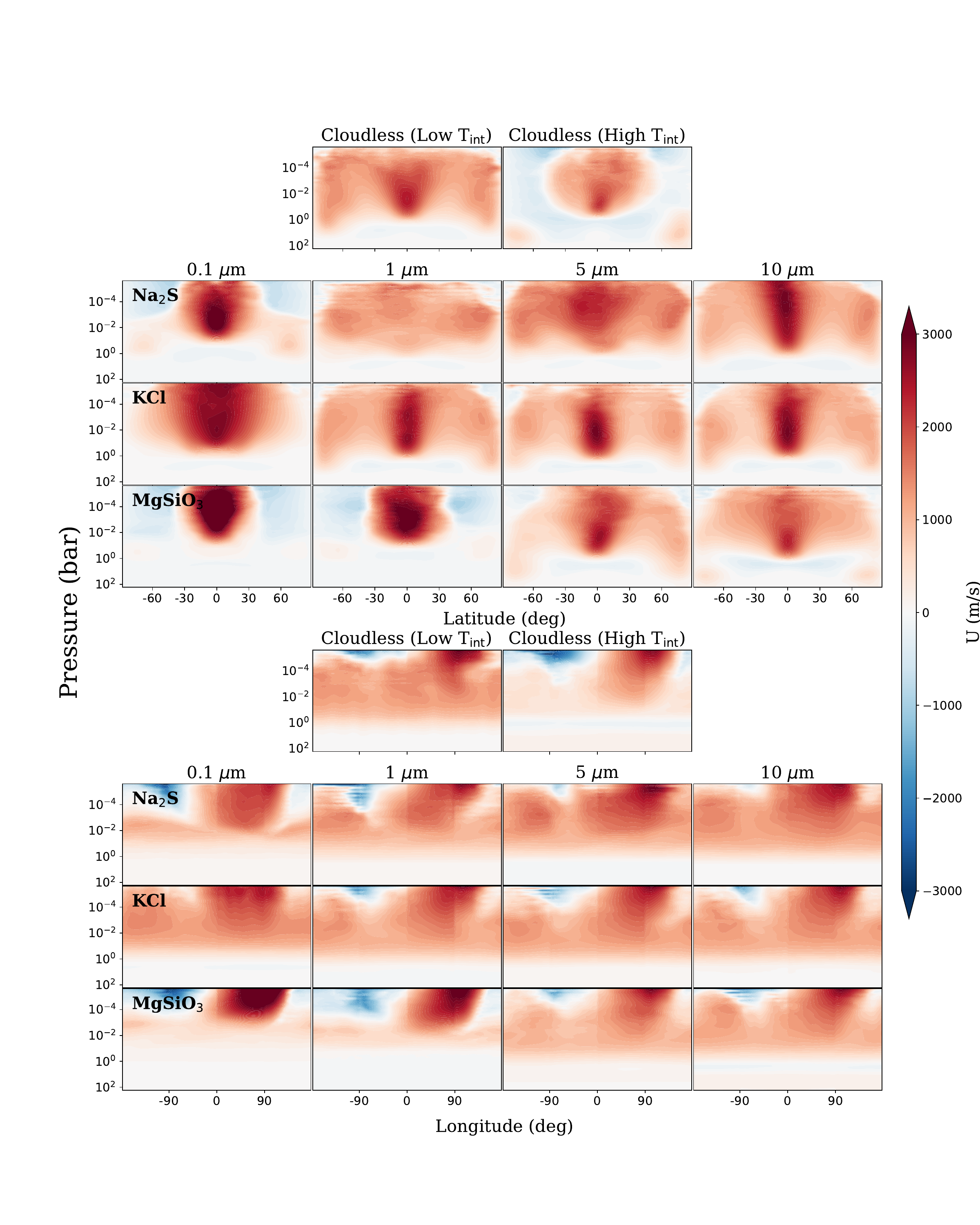}
    \caption{Same as Fig. \ref{fig:ptmean} but for zonal component of wind.}
    \label{fig:pumean}
\end{figure*}

\begin{figure*}
    \centering
    \includegraphics[width=\textwidth]{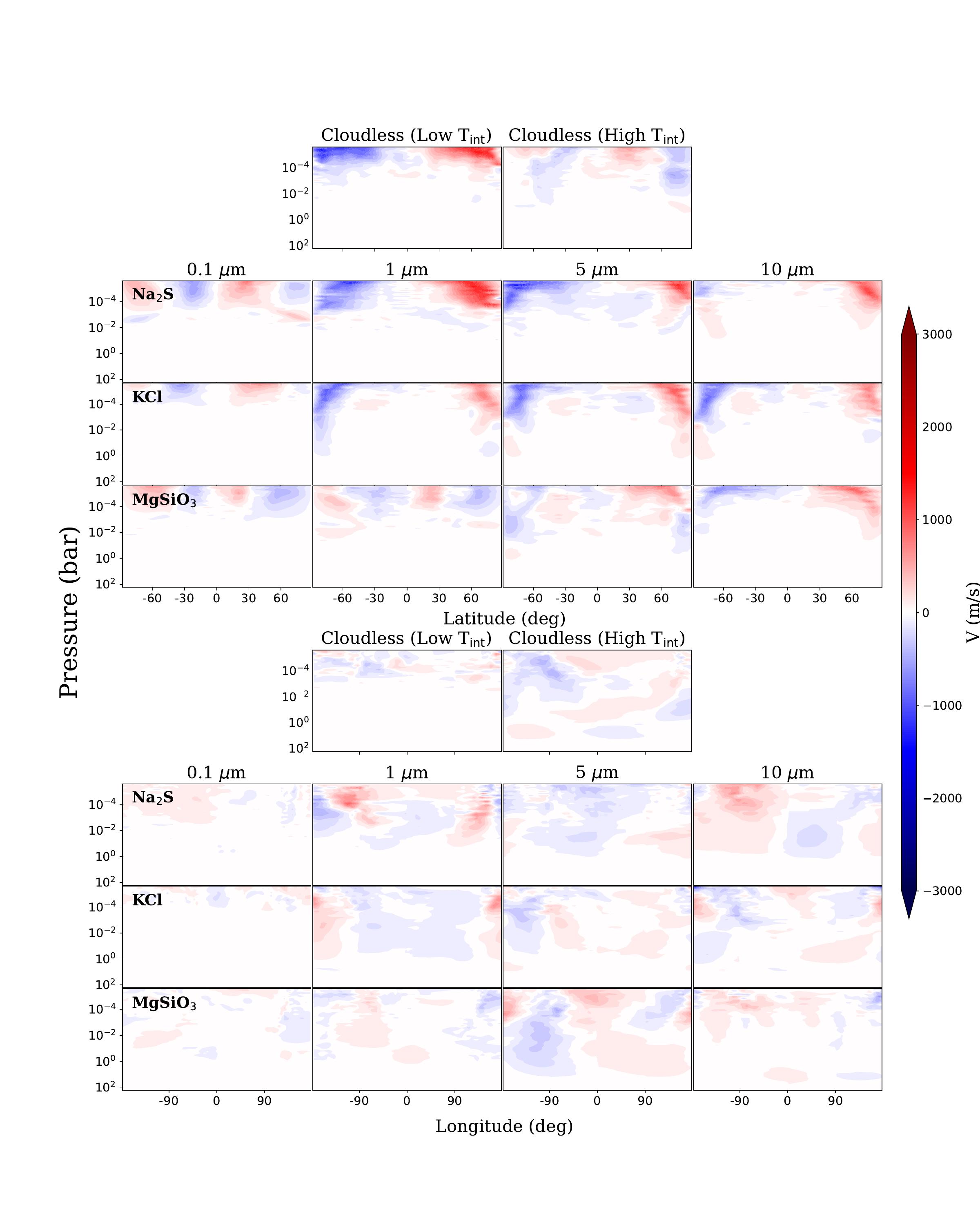}
    \caption{Same as Fig. \ref{fig:ptmean} but for meridional component of wind.}
    \label{fig:pvmean}
\end{figure*}

\begin{figure*}
    \centering
    \includegraphics[width=\textwidth]{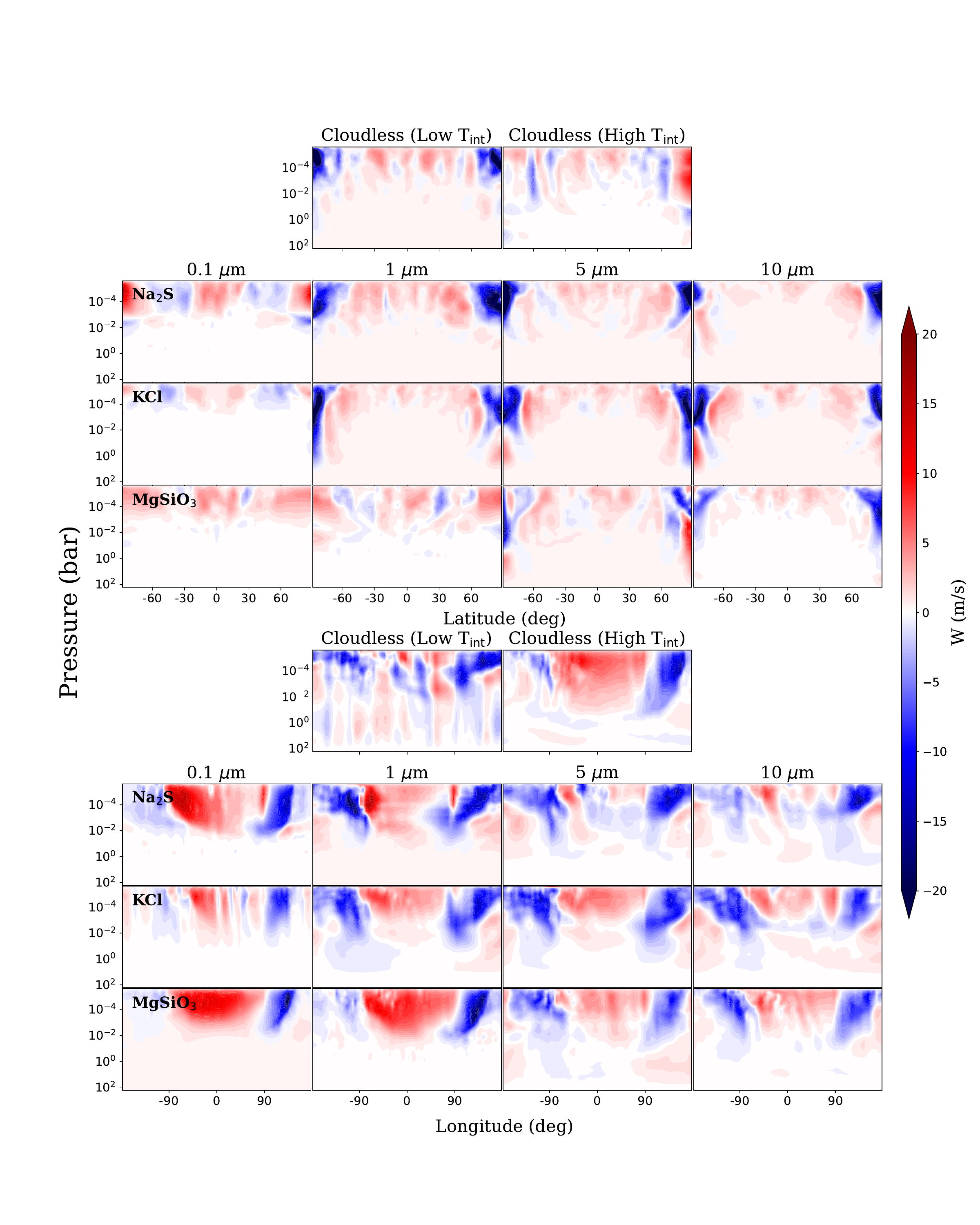}
    \caption{Same as Fig. \ref{fig:ptmean} but for vertical component of wind.}
    \label{fig:pwmean}
\end{figure*}

\begin{figure*}
    \centering
    \includegraphics[width=\textwidth]{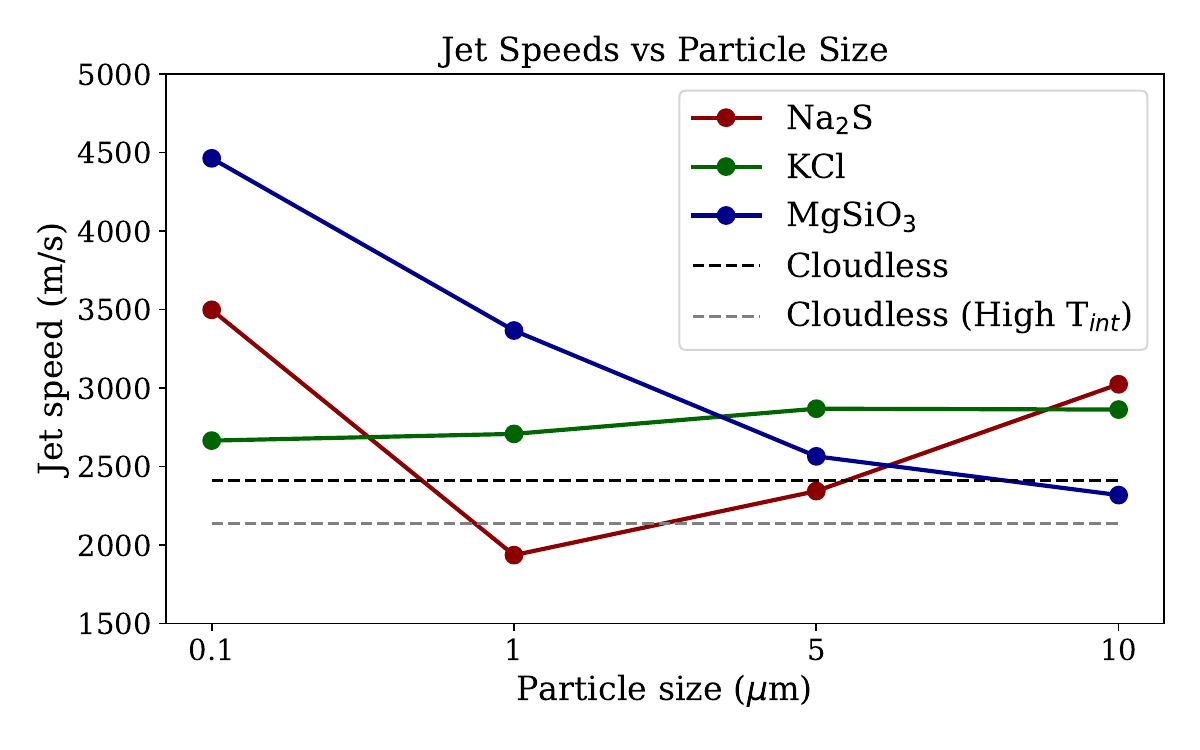}
    \caption{Jet speed (maximum zonal-mean zonal wind speed) as a function of cloud particle size, shown for each cloud species considered in the study. This plot illustrates how the characteristic particle radius influences the strength of the atmospheric jets, highlighting differences in dynamical behaviour associated with Na$_2$S, KCl, and MgSiO$_3$ clouds.}
    \label{fig:jetspeed}
\end{figure*}

\begin{figure*}
    \centering
    \includegraphics[width=\textwidth]{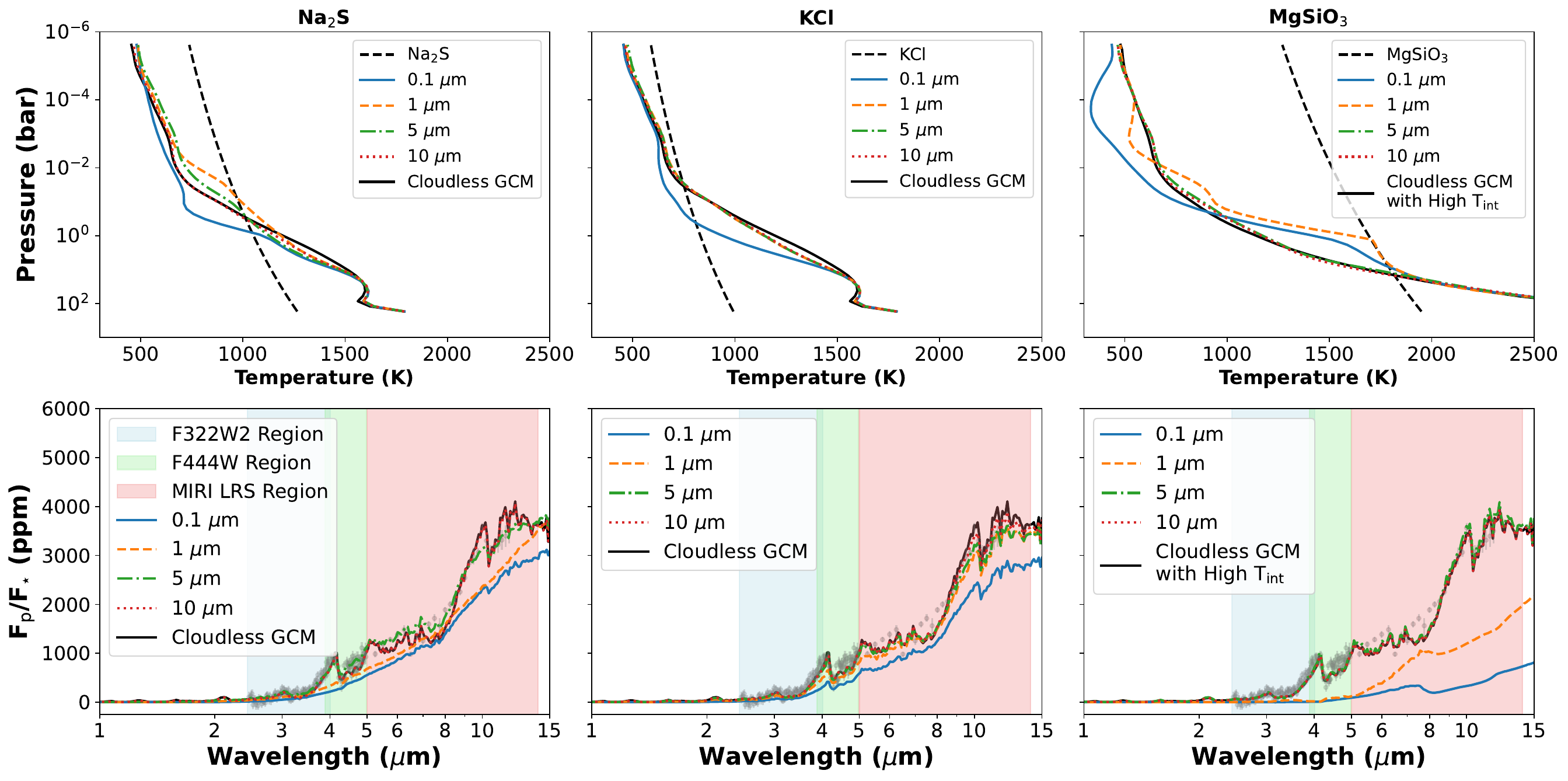}
    \caption{{Top row: Nightside averaged pressure-temperature profiles for different particle sizes along with the cloudless case (solid black line). The condensation curve for the corresponding cloud species is plotted with a black line. Bottom row: Emission spectra from GCM models along with the cloudless case (solid black line) and the dayside JWST observation (grey).}}
    \label{fig:pnight}
\end{figure*}

\end{appendix}

\end{document}